\documentclass[preprint,showpacs,preprintnumbers,amsmath,amssymb]{revtex4}

\usepackage{bm}
\usepackage[usenames]{color}
\usepackage{epsfig}
\long\def\@makecaption#1#2{%
  \par
  \vskip\abovecaptionskip
  \begingroup
   \small\rmfamily
   \sbox\@tempboxa{%
    \let\\\heading@cr
    #1 #2%
   }%
   \@ifdim{\wd\@tempboxa >\hsize}{%
    \begingroup
     \samepage
     \flushing
     \let\footnote\@footnotemark@gobble
     #1 #2\par
    \endgroup
   }{%
     \global \@minipagefalse
     \hb@xt@\hsize{\hfil\unhbox\@tempboxa\hfil}%
   }%
  \endgroup
  \vskip\belowcaptionskip
}%

\begin{document}

%This un-edited manuscript has been accepted for publication in Biophysical
%Journal and is freely available on BioFast at http://www.biophysj.org. The
%final
%copyedited version of the paper may be found at http://www.biophysj.org

\vskip 20 mm

\begin{widetext}

\noindent{\Large\bf Refolding upon force quench and pathways of mechanical and
thermal unfolding of ubiquitin}
\vskip 5 mm

\noindent{\bf Mai Suan Li$^1$, Maksim Kouza$^1$,and Chin-Kun Hu$^{2}$}
\vskip 2 mm

\noindent{\it $^1$Institute of Physics, Polish Academy of Sciences,
Al. Lotnikow 32/46, 02-668 Warszawa, Poland}

\noindent{\it $^2$Institute of Physics, Academia Sinica, Nankang,
Taipei 11529, Taiwan}

\end{widetext}

\vskip 5 mm \noindent{\bf ABSTRACT.
The refolding from stretched initial conformations of  ubiquitin
(PDB ID: 1ubq) under  the quenched force is studied using the
C$_{\alpha}$-Go model and the Langevin dynamics. It is shown that
the refolding decouples the collapse and folding kinetics. The
force quench refolding times scale as
 $\tau_F \sim \exp(f_q\Delta x_F/k_BT)$, where
 $f_q$ is the quench force and $\Delta x_F
 \approx 0.96$ nm is the location of the average transition
state along the reaction coordinate given by the end-to-end distance.
 This value is  close to $\Delta x_F \approx 0.8$ nm obtained from the
force-clamp experiments [J. M. Fernandez and H. Li, Science {\bf
303}, 1674-1678 (2004)].
The mechanical and thermal unfolding
pathways are studied and compared with the experimental and
all-atom simulation results in detail.
The sequencing of thermal
unfolding was found to be markedly different from the mechanical
one.
It is found that fixing
the N-terminus of ubiquitin changes its mechanical unfolding pathways
much more drastically compared to the case when the C-end is anchored.
We obtained the distance between the native state and the
transition state $\Delta x_{UF} \approx 0.24$ nm which is in
reasonable agreement with the experimental data.}

\vfill

Address preprint requests to Prof. Mai Suan Li, E-mail: masli@ifpan.edu.pl;
or Prof. Chin-Kun Hu, E-mail: huck@phys.sinica.edu.tw

\clearpage

\noindent
{\Large \bf Introduction}
\vskip 2 mm

\noindent

Deciphering the folding and unfolding pathways and free energy landscape
 of biomolecules remains a
challenge in molecular biology. Traditionally, folding and unfolding are
monitored by changing temperature or concentration of chemical denaturants.
In these experiments, due to thermal fluctuations of
initial unfolded conformations it is difficult to describe the folding
mechanisms in an unambiguous way. 
With the help of the atomic force microscopy, mechanical force
has been used to prepare well defined initial states of proteins
\cite{Fisher_TBS99,Fernandez_Sci04}. Using the initial force, $f_I$,
which is higher than the equilibrium critical force, $f_c$, to unfold the
tandem of poly ubiquitin (Ub), Fernandez and Li \cite{Fernandez_Sci04} have shown that
the refolding can be initiated starting from stretched conformations
or force denaturated ensemble (FDE) and quenching
the force to a low constant value, $f_q$ ($f_q < f_c$). Monitoring folding events as
a function of the end-to-end distance ($R$) they have made the following
important observations. 1) Contrary to the standard folding from the
thermal denaturated ensemble (TDE) the refolding under the quenched
force is a multiple
stepwise process and 2) The force-quench refolding time obeys the Bell formula
\cite{Bell_Sci78}, $\tau_F \approx \tau_F^0
\exp(f_q\Delta x_F/k_BT)$, where $\tau_F^0$ is the folding time
in the absence of the quench
force and $\Delta x_F$ is the average location of the transition state (TS).

Motivated by the experiments of Ferandez and Li  \cite{Fernandez_Sci04},
we have studied  \cite{Li_PNAS06}
 the refolding of the domain I27 of the human muscle protein
using the C$_{\alpha}$-Go model \cite{Clementi_JMB00}
and the four-strand
 $\beta$-barrel model sequence S1 \cite{Klimov_PNAS00}
(for this sequence the nonnative interactions are also taken into account).
Basically, we have reproduced qualitatively
the major experimental findings listed above. In addition we have
 shown that the refolding is two-state process in which the folding
 to the native basin attractor (NBA) follows
the quick collapse from initial stretched conformations with low
entropy. The corresponding kinetics can be described by the
bi-exponential time dependence, contrary to the single exponential
behavior of the folding from the TDE with high entropy.

\begin{figure}
\epsfxsize=6.3in
\centerline{\epsffile{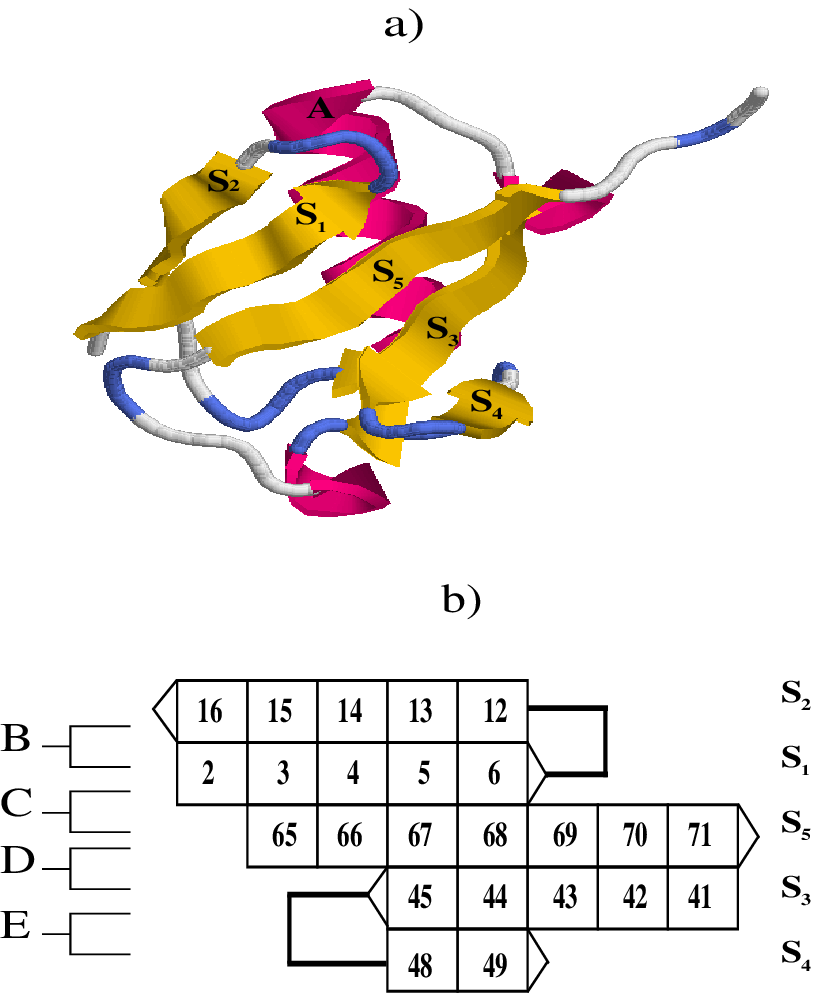}}
\caption{}
\label{ubiquitin_struture_fig}
\end{figure}

In order to make the direct comparison with the experiments of
Fernandez and Li \cite{Fernandez_Sci04}, in this paper we
performed simulations for a single domain
Ub using the C$_{\alpha}$-Go model (see
{\em Material and Method} for more details). Because the study of
refolding of 76-residue Ub (Fig. \ref{ubiquitin_struture_fig}{\em a})
by all-atom simulations is beyond
present computational facilities the Go modeling is an
appropriate choice. Most of the simulations have been carried out
at $T = 0.85T_F = 285$ K. Our present results for refolding upon
the force quench are in the qualitative agreement with the
experimental findings of Fernandez and Li, and with those obtained
for I27 and S1 theoretically \cite{Li_PNAS06}. A number of
quantitative differences between I27 and Ub will be also
discussed. For Ub we have found the average location of the
transition state $\Delta x_F \approx 0.96$ nm which is in
reasonable agreement with the experimental value 0.8 nm
\cite{Fernandez_Sci04}.

Experimentally,
the unfolding  of the polyubiquitin has been studied by applying a constant
force \cite{Schlierf_PNAS04}.
The mechanical unfolding of Ub has previously investigated using Go-like
\cite{West_BJ06} and all-atom models \cite{West_BJ06,Irback_PNAS05}.
In particular,
Irb\"ack {\em et al.} have explored mechanical
unfolding pathways 
of structures A, B, C, D and E
(see the definition of these structures and the $\beta$-strands
 in the caption to 
Fig. \ref{ubiquitin_struture_fig}) and the existence of intermediates in detail.
We present our results on mechanical unfolding of Ub
for five following reasons.
First, the barrier to the mechanical unfolding has not been computed. Second,
experiments of Schlierf {\em et al.} \cite{Schlierf_PNAS04} have suggested that
cluster 1 (strands S1, S2 and the helix A) unfolds after cluster 2
(strands S3, S4 and S5). However, this observation has not yet
been studied theoretically.
Third, since the structure C, which consists of the strands S1 and S5,
unzips first,
Irb\"ack {\em et al.} pointed out that the strand S5 unfolds before S2 or the
terminal strands follows the unfolding pathway 
S1 $\rightarrow$ S5 $\rightarrow$ S2. This conclusion may be incorrect because
it has been obtained from the breaking of the contacts within the structure C.
Fourth, in pulling and force-clamp experiments the external force is applied
to one end of proteins whereas the other end is kept fixed. Therefore, 
one important question emerges is how fixing one terminus affects the unfolding
sequencing of Ub. This issue has not been addressed by Irb\"ack {\em et al.}
\cite{Irback_PNAS05}. 
Fifth, using a simplified all-atom model it was shown \cite{Irback_PNAS05}
that mechanical intermediates occur more frequently than
in experiments \cite{Schlierf_PNAS04}. It is relevant to ask
if a C$_{\alpha}$-Go model can capture similar intermediates as this may shed
light on the role of non-native interactions.

In this paper, from the force dependence of mechanical unfolding times we
estimated the distance between the native state and the transition
state to be  $\Delta x _{UF} \approx 0.24$ nm which is close to the
experimental results of Carrion-Vazquez {\em et al.}
\cite{Carrion-Vazquez_NSB03} and Schlierf {\em et al.}
\cite{Schlierf_PNAS04}. In agreement with the experiments
\cite{Schlierf_PNAS04}, cluster 1 was found to unfold after cluster 2
in our simulations.
Applying the force to the both termini,
we studied the mechanical unfolding pathways of the terminal strands
in detail and obtained the sequencing S1 $\rightarrow$ S2 $\rightarrow$ S5
which is different from the result of Irb\"ack {\em et al.}.
When the N-terminus is fixed and the C-terminus is pulled by a
constant force the unfolding sequencing was found to be very different
from the previous case. The unzipping initiates, for example,
from the C-terminus
but not from the N-one. Anchoring the C-end is shown to have a little effect
 on unfolding pathways.
We have
demonstrated that the present C$_{\alpha}$-Go model does not capture rare
mechanical intermediates, presumably due to the lack of non-native interactions.
Nevertheless, it can correctly describe the two-state 
unfolding of Ub \cite{Schlierf_PNAS04}.

It is well known that thermal unfolding pathways may be very different
from the mechanical ones, as has been shown for the
domain I27 \cite{Paci_PNAS00}.
This is because the force is applied locally to the termini while 
thermal fluctuations have the
global effect on the entire protein. In the force case unzipping should
propagate from the termini whereas under thermal fluctuations the most
unstable part of a polypeptide chain unfolds first.

The unfolding of Ub under thermal fluctuations was investigated experimentally
by Cordier and Grzesiek \cite{Cordier_JMB02} and by Chung {\em et al.}
\cite{Chung_PNAS05}. If one assumes that unfolding is the reverse of the
 refolding process then one can infer information about the unfolding
pathways from the experimentally determined $\phi$-values \cite{Went_PEDS05}
and $\psi$-values \cite{Krantz_JMB04,Sosnick_ChemRev06}.
The most comprehensive $\phi$-value
analysis is that of Went and Jacskon. They found  that the
C-terminal region which has very low $\phi$-values unfolds first and then the
strand S1 breaks before full unfolding of the $\alpha$ helix fragment A occurs.
However, the detailed unfolding sequencing of the other strands remains unknown.
 
Theoretically, the thermal unfolding of Ub at high temperatures has been
studied by all-atom
molecular dynamics (MD) simulations by Alonso and Daggett
\cite{Alonso_ProSci98} and Larios {\em et al.} \cite{Larios_JMB04}. In
the latter work the unfolding pathways were not explored. Alonso and Daggett
have found that the $\alpha$-helix fragment A is the most resilient towards
temperature but the structure B breaks as early as the structure C.
The fact that B unfolds early contradicts not only the results for the
$\phi$-values obtained experimentally by Went and Jackson \cite{Went_PEDS05}
but also findings from a high resolution
NMR \cite{Cordier_JMB02}. Moreover, the
sequencing of unfolding events for the structures D and E was not studied.

What information about the thermal unfolding
pathways of Ub can be inferred from the folding
simulations of various coarse-grained models?  
Using a semi-empirical approach Fernandez predicted \cite{Fernandez_JCP01}
that the nucleation site involves the $\beta$-strands S1 and S5. This 
suggests that thermal fluctuations break
these strands last but
what happens to the other parts of the protein remain unknown.
Furthermore, the late breaking of S5 contradicts the unfolding
\cite{Cordier_JMB02} and folding \cite{Went_PEDS05} experiments.
From later folding simulations of Fernandez {\em et al.} 
\cite{Fernandez_Proteins02,Fernandez_PhysicaA02} one can infer
that the structures
A, B and C unzip late. Since this information is gained from $\phi$-values,
it is difficult to determine the sequencing of unfolding events even for these
fragments.
Using the results of Gilis and Rooman \cite{Gilis_Proteins01} we can
only expect that
the structures A and B unfold last. In addition,
with the help of a three-bead model it was found
\cite{Sorenson_Proteins02} that the C-terminal
loop structure is the last to fold in the folding process and most
likely plays a spectator role in the folding kinetics. This implies that
the strands S4, S5 and the second helix (residues 38-40) would unzip first
but again the full unfolding sequencing can not be inferred from this study.

Thus, neither the  direct MD \cite{Alonso_ProSci98} nor
indirect folding simulations \cite{Fernandez_JCP01,Fernandez_Proteins02,Fernandez_PhysicaA02,Gilis_Proteins01,Sorenson_Proteins02}
provide a complete picture of the thermal unfolding pathways for Ub.
One of our aims is to decipher the complete thermal unfolding sequencing
and compare
it with the mechanical one.
The mechanical and thermal routes to the denaturated states have been found
to be very different from each other.
Under the force the $\beta$-strand S1, e.g.,
unfolds first, while thermal fluctuations detach strand S5 first.
The later observation
is in good agreement with NMR data of Cordier
and Grzesiek \cite{Cordier_JMB02}.
A detailed comparison with available experimental and simulation
data on the unfolding sequencing will be presented.
The free energy barrier to thermal
unfolding was also calculated.

To summarize, in this paper we have obtained the following novel results.
We have shown that the refolding of Ub is a two-stage process in which
the "burst" phase exists on very short time scales. The construction of
the $T-f$ phase diagram allows us to determine the equilibrium critical force
$f_c$ separating the folded and unfolded regions. Using the exponential
 dependence of the refolding and unfolding
times on $f$, $\Delta x_F$ and $\Delta x_{UF}$ were computed. Our results
for $f_c$, $\Delta x_F$ and $\Delta x_{UF}$ are in acceptable agreement
with the experiments. It has been demonstrated
that fixing the N-terminus of Ub has much
stronger effect on mechanical unfolding pathways compared to the case 
when the C-end is anchored. In comparison with
previous studies, we provide a more
complete picture for thermal unfolding pathways which are very different
from the mechanical ones.

%\section{Materials and Methods}

\vskip 6 mm

\noindent{\Large \bf {Materials and Methods}} \vskip 2 mm

\noindent{\bf C$_{\alpha}$-Go model for Ub}

\vskip 2 mm

\noindent
%{\it C$_{\alpha}$-Go model for Ub}.
We use coarse-grained continuum representation for Ub
in which only the positions of C$_{\alpha}$-carbons are retained.
The interactions between residues are assumed to be Go-like and
the energy of such a model is as follows \cite{Clementi_JMB00}

\begin{eqnarray}
E \; &=& \; \sum_{bonds} K_r (r_i - r_{0i})^2 + \sum_{angles}
K_{\theta} (\theta_i - \theta_{0i})^2 \nonumber \\
&+& \sum_{dihedral} \{ K_{\phi}^{(1)} [1 - \cos (\phi_i -
\phi_{0i})] +  K_{\phi}^{(3)} [1 - \cos 3(\phi_i - \phi_{0i})] \}
\nonumber \\
& + &\sum_{i>j-3}^{NC}  \epsilon_H \left[ 5\left(
\frac{r_{0ij}}{r_{ij}} \right)^{12} - 6 \left(
\frac{r_{0ij}}{r_{ij}}\right)^{10}\right] + \sum_{i>j-3}^{NNC}
\epsilon_H \left(\frac{C}{r_{ij}}\right)^{12} - |\vec{f}.\vec{R}|
. \label{Hamiltonian}
\end{eqnarray}
Here $\Delta \phi_i=\phi_i - \phi_{0i}$,
$r_{i,i+1}$ is the distance between beads $i$ and $i+1$, $\theta_i$
is the bond angle
 between bonds $(i-1)$ and $i$,
and $\phi_i$ is the dihedral angle around the $i$th bond and
$r_{ij}$ is the distance between the $i$th and $j$th residues.
Subscripts ``0'', ``NC'' and ``NNC'' refer to the native
conformation, native contacts and non-native contacts,
respectively. Residues $i$ and $j$ are in native contact if
$r_{0ij}$ is less than a cutoff distance $d_c$ taken to be $d_c =
6.5$ \AA, where $r_{0ij}$ is the distance between the residues in
the native conformation. With this choice of $d_c$ and the native
conformation from the PDB (Fig. \ref{ubiquitin_struture_fig}{\em a}), we
have the total number of native contacts $Q_{max}=99$.

The first harmonic term in Eq. (\ref{Hamiltonian})
accounts for chain
connectivity and the second term represents the bond angle potential.
The potential for the
dihedral angle degrees of freedom is given by the third term in
Eq. (\ref{Hamiltonian}). The interaction energy between residues that are
separated by at least 3 beads is given by 10-12 Lennard-Jones potential.
A soft sphere repulsive potential
(the fourth term in Eq. \ref{Hamiltonian})
disfavors the formation of non-native contacts.
The last term accounts for the force applied to C and N termini
along the end-to-end
vector $\vec{R}$.
We choose $K_r =
100 \epsilon _H/\AA^2$, $K_{\theta} = 20 \epsilon _H/rad^2,
 K_{\phi}^{(1)} = \epsilon _H$, and
$K_{\phi}^{(3)} = 0.5 \epsilon _H$, where $\epsilon_H$ is the
characteristic hydrogen bond energy and $C = 4$ \AA. Since
$T_F=0.675 \epsilon_H$ (see below) and $T_F = 332.5 K$
\cite{Thomas_PNAS01}, we have $\epsilon_H= 4.1~{\rm kJ/mol}
= 0.98 \, {\rm kcal/mol}$. %(titin ... 1.37 Kcal/mol).
Then the force
unit $[f] = \epsilon_H/\AA \, = 68.0$ pN. %(tintin : 95.2 pN).
% Unit : 1 cal = 4.184 J

We assume the dynamics of the polypeptide chain obeys the Langevin
equation. The equations of motion (see Ref. \onlinecite{Kouza_BJ05}
for details) were integrated using the velocity form
of the Verlet algorithm \cite{Swope_JCP82}
with the time step $\Delta t = 0.005 \tau_L$,
where $\tau_L = (ma^2/\epsilon_H)^{1/2} \approx 3$ ps.

\vskip 2 mm

\noindent{\bf Simulations}

\vskip 2 mm

\noindent
%{\em Simulations }. 
In order to obtain the $T-f$ phase
diagram we use the fraction of native contacts or
the overlap function \cite{Camacho_PNAS93}
\begin{equation}
\chi \; = \frac{1}{Q_{total}} \sum_{i<j+1}^N \,\;
\theta (1.2r_{0ij} - r_{ij}) \Delta_{ij}
\label{chi_eq}
\end{equation}
where $\Delta_{ij}$ is equal to 1 if residues $i$ and $j$ form a
native contact and 0 otherwise, and $\theta (x)$ is the Heaviside
function. The argument of this function guarantees that a native
contact between $i$ and $j$ is classified as formed when $r_{ij}$
is shorter than 1.2$r_{0ij}$. The probability of being in the
native state, $f_N$, which can be measured by various experimental
techniques, is defined as $f_N = <\chi>$, where $<...>$ stands for
a thermal average. The $T-f$ phase diagram ( a plot of $1-f_N$ as
a function of $f$ and $T$) and thermodynamic quantities were
obtained by the multiple histogram method \cite{Ferrenberg_PRL89}
extended to the case when the external force is applied to the
termini \cite{Klimov_PNAS99,Klimov_JPCB01}. In this case the
reweighting is carried out not only for temperature but also for
force. We collected data for six values of $T$ at $f=0$ and for
five values of $f$ at a fixed value of $T$. The duration of MD
runs for each trajectory was chosen to be long enough to get the
system fully equilibrated (9$\times 10^5 \tau_L$ from which
1.5$\times 10^5 \tau_L$ were spent on equilibration). For a given
value of $T$ and $f$ we have generated 40 independent
trajectories for thermal averaging.

For the mechanical unfolding we have considered two cases. In the first
case the external force is applied via both termini N and C. In the second
case it is applied to either N- or C-terminus.

To simulate the mechanical unfolding the computation has been performed
at $T=285$ K and mainly at
 the constant force $f = 70, 100, 140$
and 200 pN.
%( $f$ = 0.9408, 1.344, 1.882 and 2.688 $\epsilon_H/\AA$).
This allows us to compare our results with the mechanical
unfolding experiments \cite{Schlierf_PNAS04} and to see if the
unfolding pathways change at low forces. Starting from the native
conformation but with different random number seeds the unfolding
sequencing of helix A and five $\beta$-stands is studied by
monitoring fraction of native contacts as a function of the
end-to-end extension. In the case of structures A, B, C, D and E
we consider not only the evolution of the number of
intra-structure contacts as has been done by Irb\"ack {\em et al.}
\cite{Irback_PNAS05}, but also the evolution of all contacts
(intra-structure contacts and the contacts formed by a given
structure with the rest of a protein).

In the thermal unfolding case the simulation is also started from
the native conformations and it is terminated when all of the
native contacts are broken. Due to thermal fluctuations there is
no one-to-one correspondence between $R$ and
time. Therefore $R$ ceases to be a good reaction coordinate for
describing unfolding sequencing. To rescue this, for each $i$-th
trajectory we introduce the progressive variable $\delta _i =
t/\tau^i_{UF}$, where $\tau^i_{UF}$ is the unfolding time. Then we
can average the fraction of native contacts over a unique window
$0 \le \delta _i \le 1$ and monitor the unfolding sequencing with
the  help of the progressive variable $\delta$.

%\section{Results}

\vskip 6 mm
\noindent{\Large \bf Results}
\vskip 2 mm

\noindent{\bf Temperature-force phase diagram and
thermodynamic quantities} \vskip 2 mm

\begin{figure}
\includegraphics[width=18cm,angle=0]{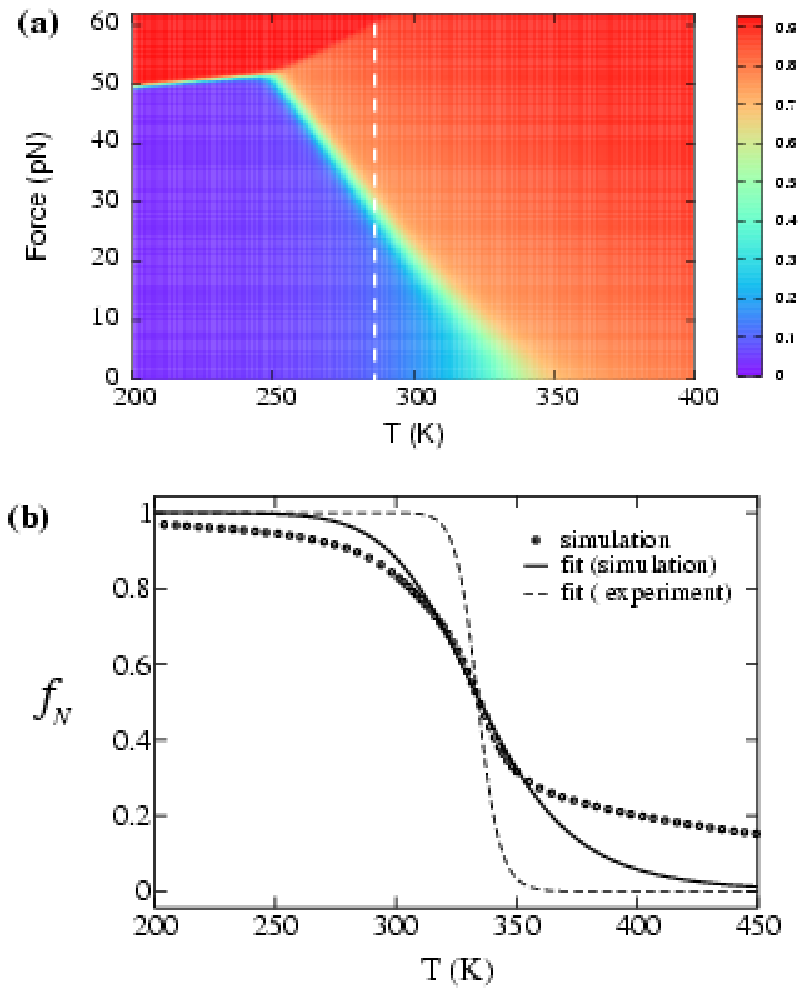}
%\vspace{0.2in}
\caption{}
\label{diagram_fN_fig}
\end{figure}

\noindent
%{\it Temperature-force phase diagram and thermodynamic
%quantities}. 
The $T-f$  phase diagram, obtained by the extended
histogram method (see {\em Materials and Methods}), is shown in
Fig. \ref{diagram_fN_fig}{\em a}. The folding-unfolding
transition, defined by the yellow region, is sharp in the low
temperature region but it becomes less cooperative (the fuzzy
transition region is wider) as $T$ increases.
The weak reentrancy (the critical force slightly increases with
$T$) occurs at low temperatures.
This seemingly strange phenomenon occurs as a result of
competition between the energy gain
and the entropy loss upon stretching.
The similar cold unzipping
transition was also observed in a number of models for
heteropolymers \cite{Shakhnovich_PRE02} and proteins
\cite{Klimov_PNAS99} including the C$_{\alpha}$-Go model for I27
(MS Li, unpublished results). As follows from the phase diagram,
at $T=285$ K the critical force $f_c \approx 30$ pN which is close
to $f_c \approx 25$ pN, estimated from the experimental pulling
data (To estimate $f_c$ from experimental pulling data
we use $f_{max} \approx f_c {\rm ln}(v/v_{min})$ \cite{Evans_BJ97},
where $f_{max}$ is the maximal force needed to unfold a protein at
the pulling speed $v$. From the raw data in Fig. 3b of
Ref. \cite{Carrion-Vazquez_NSB03} we obtain $f_c \approx$ 25 pN).
%\cite{Li_fit}.  
Given the simplicity
of the model this agreement can be considered satisfactory and it validates
the use of the Go model.

Figure \ref{diagram_fN_fig}{\em b} shows the temperature
dependence of population of the native state $f_N$. Fitting to the
standard two-state curve $f_N = \frac{1}{1 + \exp[-\Delta
H_m(1-\frac{T}{T_m})/k_BT]}$ one can see that it works pretty well
(solid curve) around the transition temperature but it gets worse
at high $T$ due to slow decay of $f_N$. Such a behavior is
characteristic for almost all of theoretical models
\cite{Kouza_BJ05} including the all-atom ones
\cite{Phuong_Proteins05}. In fitting we have chosen the hydrogen
bond energy $\epsilon_H = 0.98$ kcal/mol in Hamiltonian
(\ref{Hamiltonian}) so that $T_F = T_m = 0.675 \epsilon_H/k_B$
coincides with the experimental value 332.5 K
\cite{Thomas_PNAS01}. From the fit we obtain $\Delta H_{\rm m} =
11.4$ kcal/mol which is smaller than the experimental value 48.96
kcal/mol indicating that the Go model is, as expected, less stable
compared to the real Ub. Taking into account non-native contacts
and more realistic interactions between side chain atoms is
expected to increase the stability of the system.

The cooperativity of the denaturation transition may be characterized by
the cooperativity index, $\Omega_c$ (see Refs. \onlinecite{Klimov_FD98}
and \onlinecite{Li_PRL04}
for definition). From simulation data for $f_N$ presented in
Fig. \ref{diagram_fN_fig}{\em b} we have
 $\Omega_c \approx 57$ which is
considerably lower than the experimental value $\Omega_c \approx
384$ obtained with  the help of $\Delta H_{\rm m}$ = 48.96
kcal/mol and $T_m = 332.5$K \cite{Thomas_PNAS01} . The
underestimation of $\Omega _c$ in our simulations is not only a
shortcoming of the off-lattice Go model \cite{Kouza_JPCA06} but
also a common problem of much more sophisticated force fields in
all-atom models \cite{Phuong_Proteins05}.

Another measure of the cooperativity is the ratio between the
van't Hoff and the calorimetric enthalpy $\kappa _2$
\cite{Kaya_PRL00}. For the Go Ub we obtained $\kappa _2 \approx
0.19$. Applying the base line subtraction \cite{Chan_ME04} gives
$\kappa _2 \approx 0.42$ which is still much below $\kappa _2
\approx 1 $ for the truly one-or-none transition. Since $\kappa
_2$ is an extensive parameter, its low value is due to the
shortcomings of the off-lattice Go models but not due to the
finite size effects. More rigid lattice models give better results
for the calorimetric cooperativity $\kappa _2$
\cite{Li_Physica05}.

\begin{figure}
\includegraphics[width=17cm,angle=0]{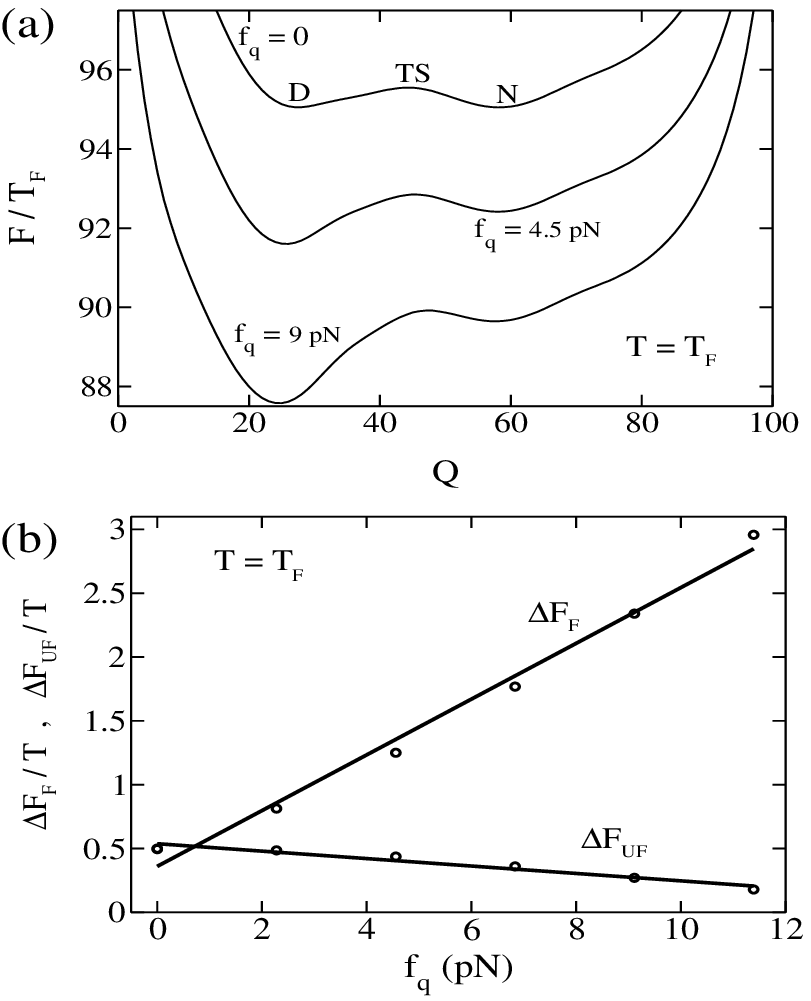}
\vspace{0.2in}
 \caption{}
\label{free_Q_barrier_fig}
\end{figure}

Figure \ref{free_Q_barrier_fig}{\em a} shows the free energy as a
function of $Q$ for several values of force at $T=T_F$. Since
there are only two minima, our results support the two-state
picture of Ub \cite{Schlierf_PNAS04,Chung_PNAS05}. As expected,
the external
 force increases the folding barrier, $\Delta F_F$
($\Delta F_F = F_{TS} - F_D$) and it lowers
the unfolding barrier, $\Delta F_{UF}$ ($\Delta F_{UF} = F_{TS} - F_N$).
From the linear fits in
Fig. \ref{free_Q_barrier_fig}{\em b} we obtain the average distance between the
TS and D states, $\Delta x_F = \Delta F_F/f \approx 1$ nm,
and the distance between TS and the native state,
$\Delta x_{UF} = \Delta F_{UF}/f \approx 0.13$ nm.
Note that $\Delta x_F$ is very
close to $\Delta x_F \approx$ 0.96 nm obtained from refolding
times at a bit lower temperature $T=285$ K (see Fig. \ref{Kf_vs_ffc_fig} below).
However, $\Delta x_{UF}$ is lower than value 0.24 nm followed
from mechanical unfolding data at $f > f_c$
(Fig. \ref{uftime_low_high_fig}). This difference may be caused
by either sensitivity of $\Delta x_{UF}$ to the temperature or the determination
of $\Delta x_{UF}$ from the
approximate free energy landscape as a function of a single
coordinate $Q$ is not sufficiently accurate.

We have also studied the free energy landscape using $R$ as a reaction
coordinate.
The dependence of $F$ on $R$ was found to be smoother
(results not shown) compared
to what was obtained by Kirmizialtin {\em et al.} \cite{Kirmizialtin_JCP05}
using a more elaborated model
\cite{Sorenson_Proteins02}
which involves the non-native interactions.

%{\em Refolding under quenched force.}

\vskip 2 mm

\noindent{\bf Refolding under quenched force} \vskip 2 mm

\noindent
Our protocol for studying the refolding of Ub is identical to what has been
done on the experiments of Fernandez and Li \cite{Fernandez_Sci04}.
We first apply the force $f_I \approx 70$ pN to prepare initial
conformations (the
protein is stretched if $R \ge 0.8 L$, where the contour length $L = 28.7 $ nm).
Starting from the FDE we quenched the force
to $f_q < f_c$ and then monitored the refolding process by following
the time dependence of the number of native
contacts $Q(t)$, $R(t)$ and the radius of gyration
$R_g(t)$ for typically 50 independent trajectories.

\begin{figure}
\includegraphics[width=16cm,angle=0]{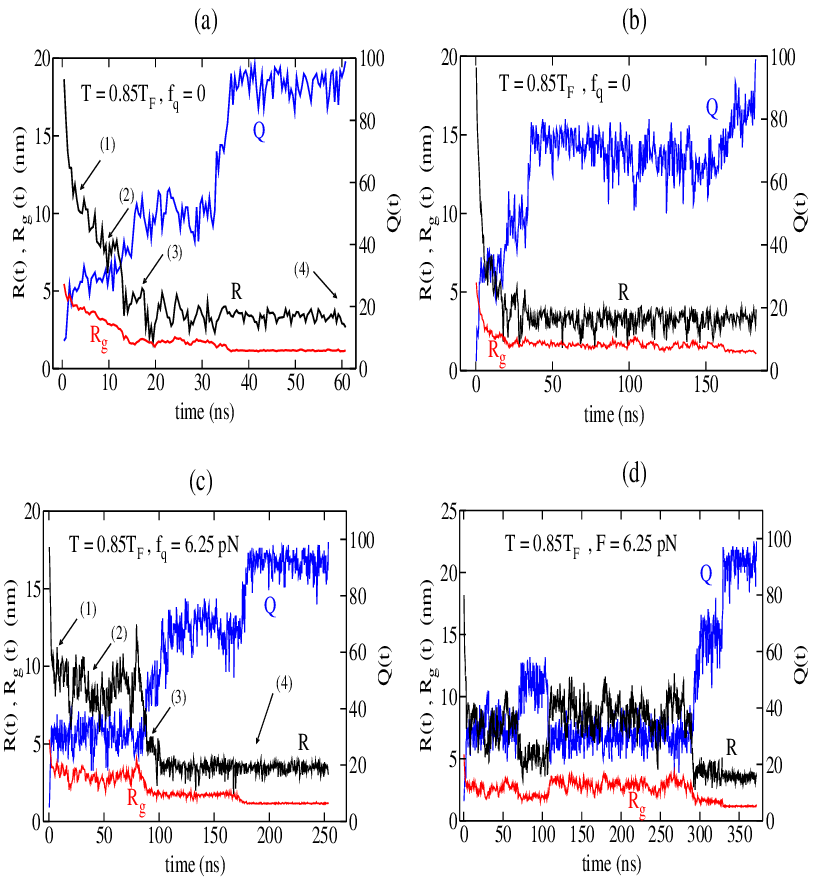}
%\vspace{0.2in}
\caption{}
\label{ContRgR_time_fig}
\end{figure}

Figure  \ref{ContRgR_time_fig} shows considerable diversity of
refolding pathways. In accord with experiments
\cite{Fernandez_Sci04} and simulations for I27 \cite{Li_PNAS06},
the reduction of $R$ occurs in a stepwise manner. In the $f_q=0$
case (Fig. \ref{ContRgR_time_fig}{\em a}) $R$ decreases
continuously from $\approx 18$ nm to 7.5 nm (stage 1) and
fluctuates around this value for about 3 ns (stage 2). The further
reduction to $R \approx 4.5$ nm (stage 3) until a transition to
the NBA. The stepwise nature of variation of $Q(t)$ is also
clearly shown up but it is more masked for $R_g(t)$. Although we
can interpret another trajectory for $f_q=0$ (Fig.
\ref{ContRgR_time_fig}b) in the same way, the time scales are
different. Thus, the refolding routes are highly heterogeneous.

The pathway diversity is also evident for $f_q >0$
(Fig. \ref{ContRgR_time_fig}{\em c}
and {\em d}). Although the picture remains qualitatively the same as in the
$f_q=0$ case, the time scales for different steps becomes much larger.
The molecule fluctuates around $R \approx 7$ nm, e.g.,
for $\approx 60$ ns (stage 2 in Fig. \ref{ContRgR_time_fig}{\em c})
which is considerably longer
than $\approx 3$ ns in Fig. \ref{ContRgR_time_fig}{\em a}.
The variation of $R_g(t)$ becomes more drastic
compared to the $f_q=0$ case.

\begin{figure}
\includegraphics[width=16cm,angle=0]{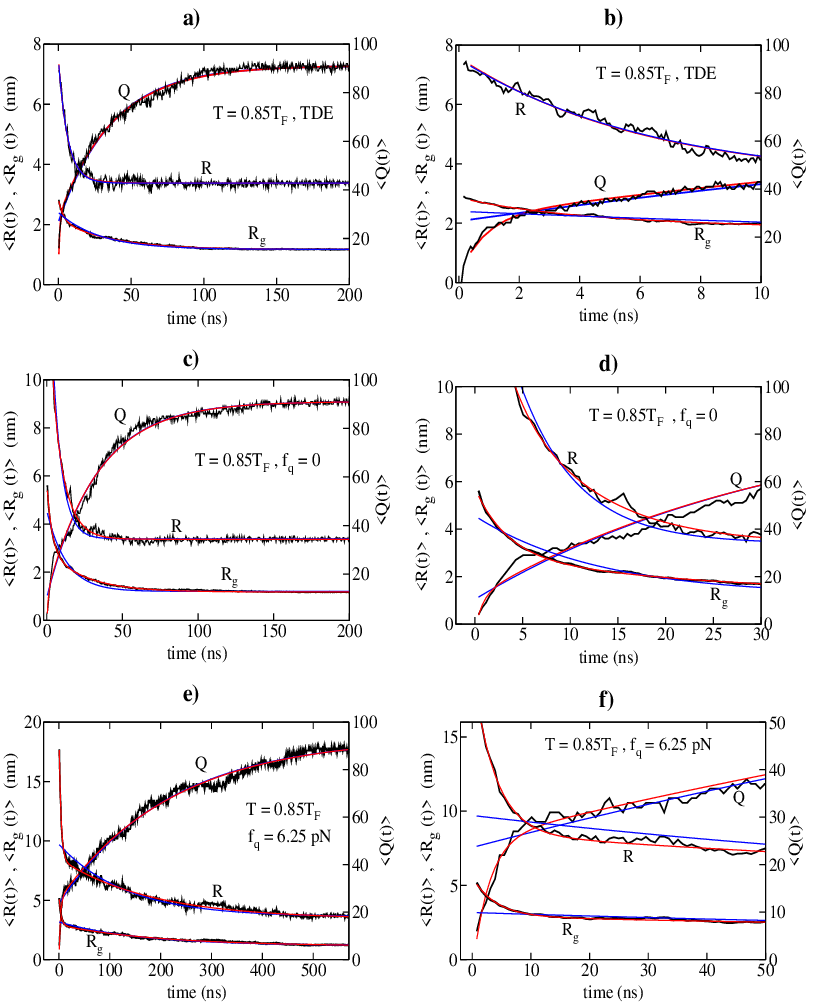}
%\vspace{0.2in}
\caption{}
\label{ContRgR_time_av_fig}
\end{figure}

Figure \ref{ContRgR_time_av_fig} shows the time dependence of
$<R(t)>, <Q(t)>$ and $<R_g(t)>$, where $<...>$ stands for
averaging over 50 trajectories. The left and right panels
correspond to the long and short time windows, respectively. For
the TDE case (Fig. \ref{ContRgR_time_av_fig}{\em a} and {\em b})
the single exponential fit works pretty well for $<R(t)>$ for the
whole time interval. A little departure from this behavior is seen
for $<Q(t)>$ and $<R_g(t)>$ for $t < 2$ ns (Fig.
\ref{ContRgR_time_av_fig}{\em b}). Contrary to the TDE case, even
for $f_q=0$ (Fig. \ref{ContRgR_time_av_fig}{\em c} and {\em d})
the difference between the single and bi-exponential fits is
evident not only for $<Q(t)>$ and $<R_g(t)>$  but also for
$<R(t)>$. The time scales, above which two fits become eventually
identical, are slightly different for three quantities (Fig.
\ref{ContRgR_time_av_fig}{\em d}). The failure of the single
exponential behavior becomes more and more evident with the
increase of $f_q$, as demonstrated in Figs.
\ref{ContRgR_time_av_fig}{\em e} and {\em f} for the FDE case with
$f_q = 6.25$ pN.

Thus, in agreement with our previous results, obtained for I27 and
the sequence S1 \cite{Li_PNAS06},
starting from FDE the refolding kinetics compiles of the fast and
slow phase. The characteristic time scales for these phases may be
obtained using a sum of two exponentials,$<A(t)> = A_0 + A_1
\exp(-t/\tau^A_1) + A_2 \exp(-t/\tau^A_2)$, where $A$ stands for
$R$, $R_g$ or $Q$. Here $\tau^A_1$ characterizes the burst-phase
(first stage) while $\tau^A_2$ may be either the collapse time
(for $R$ and $R_g$) or the folding time (for $Q$) ($\tau^A_1 <
 \tau^A_2$). As in the case of I27 and S1 \cite{Li_PNAS06},
$\tau^R_1$ and $\tau^{R_g}_1$ are almost independent on $f_q$
(results not shown). We attribute this to the fact that the quench
force ($f_q^{max} \approx 9$ pN) is much lower than the entropy
force ($f_e$) needed to stretch the protein. At $T=285$ K, one has
to apply $f_e \approx 140$ pN for stretching Ub to 0.8 $L$. Since
$f_q^{max} << f_e$ the initial compaction of the chain that is
driven by $f_e$ is not sensitive to the small values of $f_q$.
Contrary to $\tau^A_1$, $\tau^A_2$ was found to increase with $f_q$
exponentially. Moreover,
$\tau^R_2 < \tau^{R_g}_2 < \tau _F$ implying that the chain compaction
occurs before the acquisition of the native state.

Figure \ref{Kf_vs_ffc_fig} shows the dependence of the folding
times on $f_q$. Using the Bell-type formula \cite{Bell_Sci78} and
the linear fit  in Fig. \ref{Kf_vs_ffc_fig} we obtain $\Delta x_F
\approx 0.96$ nm which is in acceptable agreement with the
experimental
 value $\Delta x_F \approx 0.8$ nm
\cite{Fernandez_Sci04}.
The linear growth of the free energy barrier to folding with $f_q$
is due to
the stabilization of the random coil states under the force.
 Our estimate for Ub is higher than
$\Delta x_F \approx 0.6$ nm obtained for I27 \cite{Li_PNAS06}.
One of possible reasons for such a pronounced difference is that we used
the cutoff distance $d_c=0.65$ and 0.6 nm in the Go model (\ref{Hamiltonian})
for Ub and I27, respectively. The larger value of $d_c$ would make a protein
more stable (more native contacts) and it may change the free energy landscape
leading to enhancement of $\Delta x_F$. This problem requires
further investigation.

%{\em Absence of mechanical unfolding intermediates in
%C$_{\alpha}$-Go model.}

\vskip 2 mm

\noindent{\bf Absence of mechanical unfolding intermediates in
C$_{\alpha}$-Go model} \vskip 2 mm

\noindent
 In order to study the unfolding dynamics
of Ub,  Schlierf {\em et al.} \cite{Schlierf_PNAS04} have
performed the AFM experiments at a constant force $f = 100, 140$
and 200 pN.  The unfolding intermediates were recorded in about $5
\%$ of 800 events at different forces. The typical distance
between the initial and intermediate states is $\Delta R = 8.1 \pm
0.7$ nm \cite{Schlierf_PNAS04}. However, the intermediates do not
affect the two-state behavior of the polypeptide chain. Using the
all-atom models Irb\"ack {\em et al.} \cite{Irback_PNAS05} have also
observed the intermediates in the region 6.7 nm $< R < 18.5$ nm.
Although the percentage of intermediates is higher than in the
experiments, the two-state unfolding events remain dominating.
To check the existence of force-induced intermediates in our model, we
have performed the unfolding simulations for $f=70, 100,
140$ and 200 pN. Because the results are qualitatively similar for
all values of force, we present $f=100$ pN case only.

Figure \ref{uf100pN_long_fig} shows the time dependence of $R(t)$
for fifteen runs starting from the native value $R_N \approx 3.9$ nm.
For all trajectories the plateau occurs at $R \approx 4.4$ nm. As
seen below, passing this plateau corresponds to breaking of intra-structure
native contacts of structure C. At this stage the chain ends get almost
stretched out, but the rest of the polypeptide chain remains
native-like. The plateau is washed out when we average over many
trajectories and $<R(t)>$ is well fitted by a single exponential
 (Fig. \ref{uf100pN_long_fig}), in accord
with the two-state behavior of Ub \cite{Schlierf_PNAS04}.

The existence of the plateau observed for individual unfolding
events in Fig. \ref{uf100pN_long_fig} agrees with the all-atom
simulation results of Irb\"ack {\em et al.} \cite{Irback_PNAS05} who
have also recorded the similar plateau at $R \approx 4.6$ nm at
short time scales. However unfolding intermediates at larger
extensions do not occur in our simulations. This is probably
related to neglect  of the non-native interactions in the
C$_{\alpha}$-Go model. Nevertheless, this simple model provides
the correct two-state unfolding picture of Ub in the statistical
sense.

%{\em Mechanical unfolding barrier.}

\vskip 2 mm
\noindent{\bf Mechanical unfolding barrier} \vskip 2 mm

\noindent
We now try to determine the barrier
to the mechanical unfolding from the dependence of the unfolding
times $\tau _{UF}$ on $f$.
It should be noted that this way of determination of the unfolding barrier
is exact and it would give a more reliable estimate compared to the free
energy landscape approach in which the free energy profile is approximated
as a function of only one order parameter.

We first consider the case when the force is applied via both termini N and C.
Since
the force lowers the unfolding barrier, $\tau _{UF}$ should
decrease as 
$f$ increases (Fig. \ref{uftime_low_high_fig}). The present Go model
gives $\tau _{UF}$ smaller than the experimental values by about
eight orders of magnitude. E.g., for $f=100$ pN, $\tau _{UF}
\approx 12$ ns whereas the experiments gives $\tau_{UF} \approx$
2.77 s \cite{Schlierf_PNAS04}. As seen from Fig.
\ref{uftime_low_high_fig}, for $f < 140$ pN $\tau_{UF}$ depends on
$f$ exponentially. In this regime $\tau _{UF} \approx \tau_{UF}^0
\exp(fx_{UF}/k_BT)$, where $\Delta x_{UF}$
 is the average
distance between the N and TS states. From the linear fit in Fig.
\ref{uftime_low_high_fig} we obtained $\Delta x_{UF} \approx$ 0.24
nm. Using different fitting procedures Schlierf {\em et al.}
\cite{Schlierf_PNAS04} have obtained $\Delta x_{UF} \approx$ 0.14
nm and 0.17 nm. The larger value $\Delta x_{UF} \approx$ 0.25 nm
was reported in the earlier experiments
\cite{Carrion-Vazquez_NSB03}. Thus, given experimental
uncertainty, the C$_{\alpha}$-Go model provides a reasonable
estimate of $\Delta x_{UF}$ for the two-state Ub.

In the high force regime ($f > 140$ pN) instead of the exponential
dependence $\tau_{UF}$ scales with $f$ linearly (inset in Fig.
\ref{uftime_low_high_fig}). The crossover from the exponential to
the linear behavior is in full agreement with the earlier
theoretical prediction \cite{Evans_BJ97}. The similar crossover has been
also observed \cite{Szymczak_JPCM06} for the another Go-like model
of Ub but $\Delta x_{UF}$ has not been estimated. At very high
forces $\tau_{UF}$ is expected to be asymptotically independent of
$f$.

One can show that fixing one terminus of a protein has the same effect on
unfolding times no matter  the N- or C-terminus is fixed.
Therefore, we show the
results obtained for the case when the N-end is anchored. As seen from
Fig. \ref{uftime_low_high_fig}, 
the unfolding process is slowed down nearly by a factor of 2.
It may imply that diffusion-collision processes
\cite{Karplus_Nature76} play an important role in
the Ub unfolding. Namely, as follows from the diffusion-collision model,
the time, required for formation (breaking) contacts, is inversely
proportional to the diffusion coefficient, $D$, of a pair of spherical
units. If one of them is idle, $D$ is halved and
the time needed to break contacts increases accordingly.  
Although fixing one end increases the unfolding times,
it does not change the distance between the TS and the native state,
$\Delta x_{UF}$ (Fig. \ref{uftime_low_high_fig}).

%{\em Mechanical unfolding pathways.}

\vskip 2 mm

\noindent{\bf Mechanical unfolding pathways: force is applied to both termini} \vskip 2 mm

\noindent
Here we focus on the
mechanical unfolding pathways by monitoring the number of native
contacts as a function of the end-to-end extension $\Delta R
\equiv R-R_{\rm eq}$, where $R_{\rm eq}$ is the
equilibrium value of $R$. For $T=285$ K $R_{\rm eq} \approx 3.4$ nm.
Following Schlierf {\em et al.}
\cite{Schlierf_PNAS04}, we first divide Ub into two clusters.
Cluster 1 consists of strands S1, S2 and the helix A (42 native
contacts) and cluster 2 - strands S3, S4 and S5 (35 native
contacts). The dependence of fraction of intra-cluster native
contacts is shown in Fig. \ref{clus_intra_ext_fig} for $f = 70$
and 200 pN (similar results for $f = 100$  and 140 pN are not shown). In
agreement with the experiments \cite{Schlierf_PNAS04} the cluster
2 unfolds first. The unfolding of these clusters becomes more and
more synchronous upon decreasing $f$. At $f = 70$ pN the competition
with thermal
fluctuations becomes so important that two clusters may unzip
almost simultaneously. Experiments at low forces are needed to
verify this observation.

The arrow in Fig. \ref{clus_intra_ext_fig} marks the position
$\Delta R = 8.1$ nm,
where some intermediates were recorded in the experiments
\cite{Schlierf_PNAS04}. At this point
there is intensive loss of native contacts of the cluster 2 suggesting that
the intermediates observed on the experiments are conformations in which
most of the contacts of this
cluster are already broken but the cluster 1 remains relatively
structured ($\approx 40\%$ contacts). One can expect that the cluster 1
is more ordered in the intermediate conformations if the side chains and
realistic interactions between amino acids are taken into account.

To compare the mechanical unfolding pathways of Ub with the
all-atom simulation results \cite{Irback_PNAS05} we discuss the
sequencing of helix A and structures B, C, D and E in more detail. We
monitor the intra-structure native contacts and all contacts
separately. The later include not only the contacts within a given
structure but also the contacts between it and the rest of the
protein. It should be noted that Irb\"ack {\em et al.} have studied
the unfolding pathways based on the evolution of the
intra-structure contacts. Fig. \ref{dom_ext_100pN_fig}a shows the
dependence of the fraction of intra-structure contacts on $\Delta
R$ at $f=100$ pN. At $\Delta R \approx $ 1nm, which corresponds to
the plateau in Fig. \ref{uf100pN_long_fig}, most of the contacts
of C are broken. In agreement with the all-atom simulations
\cite{Irback_PNAS05}, the unzipping follows C $\rightarrow$ B
$\rightarrow$ D $\rightarrow$ E $\rightarrow$ A. Since C consists
of the terminal strands S1 and S5, it was suggested that these
fragments unfold first. However, this scenario may be no
longer valid if one considers not only intra-structure
contacts but also other possible ones (Fig.
\ref{dom_ext_100pN_fig}{\em b}). In this case the statistically
preferred sequencing is B $\rightarrow$ C $\rightarrow$ D
$\rightarrow$ E $\rightarrow$ A which holds not only for $f$=100
pN but also for other values of $f$. If it is true then S2 unfold
even before S5. To make this point more transparent,
 we plot the fraction of contacts for S1, S2 and S5 as a
 function of $\Delta R$ (Fig. \ref{str_detach_cont_tr75_200pn_fig}{\em a})
for a typical trajectory.
Clearly, S5 detaches from the core part of a protein
after S2 (see also the snapshot
in Fig. \ref{str_detach_cont_tr75_200pn_fig}{\em b}).
So, instead of the sequencing S1 $\rightarrow$ S5 $\rightarrow$ S2
proposed by Irb\"ack {\em et al.}, we obtain
S1 $\rightarrow$ S2 $\rightarrow$ S5.

The dependence of the fraction
of native contacts on $\Delta R$ for individual strands
is shown in Fig. \ref{cont_ext_fig}{\em a} ($f=70$ pN) and
Fig. \ref{cont_ext_fig}{\em b}
($f$=200 pN). At $\Delta = 8.1$ nm contacts of S1, S2 and S5 are already broken
whereas S4 and A remain largely structured. In terms of $\beta$-strands and A
we can interpret the intermediates observed in the experiments of
Schlierf {\em et al.} \cite{Schlierf_PNAS04} as conformations with
well structured S4 and A, and low ordering of S3. This interpretation is
more precise compared to the above argument
based on unfolding of two clusters because if one considers the average
number of native contacts, then
the cluster 2 is unstructured in the intermediate state
(Fig. \ref{clus_intra_ext_fig}), but its strand S4 remains highly structured
(Fig. \ref{cont_ext_fig}).

From Fig. \ref{cont_ext_fig} we obtain the following mechanical unfolding
sequencing
\begin{equation}
{\rm S1} \rightarrow {\rm S2} \rightarrow {\rm S5} 
 \rightarrow {\rm S3} \rightarrow {\rm S4} \rightarrow {\rm A}.
\label{mechanical_sequencing}
\end{equation}
It should be noted that the sequencing
(\ref{mechanical_sequencing}) is valid in the statistical  sense.
In some trajectories S5 unfolds even before S1 and S2 or the
native contacts of S1, S2 and S5 may be broken at the same time
scale (Table 1). From the Table 1
it follows that the probability of having
S1 unfolded first decreases with lowering $f$ but the main
trend (\ref{mechanical_sequencing}) remains unchanged.
One has to stress again that the sequencing of the terminal strands
S1, S2 and S5 given by Eq. \ref{mechanical_sequencing} is different from
what proposed by Irb\"ack {\em et al.} based on the breaking of 
the intra-structure contacts of C.
Unfortunately, there are no experimental data available for
comparison with our theoretical prediction.

\vskip 2 mm

\noindent{\bf Mechanical unfolding pathways: One end is fixed} \vskip 2 mm

\noindent
{\em N-terminus is fixed}.
Here we adopted the same procedure as in the previous section
except the N-terminus is held fixed during simulations. As in the process
where both of the termini are subjected to force, one can show that
the cluster 1
unfolds after the cluster 2 (results not shown).

 From Fig. \ref{cont_snap_fixN_f200pN_fig}
we obtain the following unfolding pathways
\begin{subequations}
\begin{equation}
{\rm C} \rightarrow {\rm D} \rightarrow {\rm E} \rightarrow {\rm B} \rightarrow {\rm A},
\label{mechan_fixN_sequencing_struc}
\end{equation}
\begin{equation}
{\rm S5} \rightarrow {\rm S3} \rightarrow {\rm S4} \rightarrow {\rm S1} \rightarrow {\rm S2}
 \rightarrow {\rm A},
\label{mechan_fixN_sequencing}
\end{equation}
\end{subequations}
which are also valid for the other values of force ($f$=70, 100 and 140 pN).
Similar to the case when the force is applied to both ends, the structure
C unravels first and the helix A remains the most stable. However, the
sequencing of B, D and E changes markedly compared to the result
obtained by Irb\"ack {\em et al} \cite{Irback_PNAS05} 
(Fig. \ref{dom_ext_100pN_fig}a).

As evident from Eqs. \ref{mechanical_sequencing} and \ref{mechan_fixN_sequencing}, anchoring the first terminal has a much more pronounced effect on the 
unfolding pathways of individual strands. In particular,
unzipping commences from the
C-terminus instead of from the N-one.
Fig. \ref{cont_snap_fixN_f200pN_fig}{\em c} shows a typical snapshot
where one can see clearly that S$_5$ detaches first. At the first glance,
this fact may seem trivial because S$_5$ experiences the external force
directly.
However,  our experience on unfolding pathways of the well studied domain I27
from the human cardiac titin, e.g., shows that it may be not the case.
Namely, as follows from the pulling experiments \cite{Marszalek_Nature99}
and simulations \cite{Lu_Proteins99},
the strand A from the N-terminus 
 unravels first although this terminus
is kept fixed. From this point of view, what strand of Ub detaches first
is not {\em a priori} clear. In our opinion, it depends on the
interplay between the native topology and the speed of tension propagation.
The later factor probably plays a more important role for Ub while the
opposite situation happens with I27.
One of possible reasons is related to the high stability of the helix A
which does not allow either for the N-terminal to unravel first or
for seriality in unfolding starting from the C-end. 

{\em C-terminus is fixed}.
One can show that unfolding pathways of structures A,B, C, D and E remain
exactly the same as in the case when Ub has been pulled from both termini
(see Fig. \ref{dom_ext_100pN_fig}). Concerning the individual strands,
a slight difference is observed for S$_5$ 
(compare Fig. \ref{cont_snap_fixN_f200pN_fig}{\em d} and
Fig. \ref{cont_ext_fig}). Most of the native contacts of this domain break
before S$_3$ and S$_4$, except
the long tail at extension $\Delta R \gtrsim$ 11 nm due to high
mechanical stability of
only one
contact between residues 61 and 65 (the highest resistance of
this pair is probably due to the fact that among 25 possible contacts
of S$_5$ it has the shortest distance $|61-65|=4$ in sequence).
This scenario holds in about 90\%
of trajectories whereas
S$_5$ unravels 
completely earlier than S$_3$ and S$_4$ in the remaining trajectories.
Thus, anchoring C-terminus has much less effect on unfolding pathways compared
to the case when the N-end is immobile.

It is worth to note that, experimentally one has studied the effect of
 extension geometry on the 
mechanical stability of Ub fixing its C-terminus \cite{Carrion-Vazquez_NSB03}.
The greatest mechanical strength (the longest unfolding time) occurs when
the protein is extended between N- and C-termini. This result has been
supported by Monte Carlo \cite{Carrion-Vazquez_NSB03} as well as MD
\cite{West_BJ06} simulations. However the mechanical
unfolding sequencing has not
been studied yet. It would be interesting to check our results on the
effect of fixing one end on Ub mechanical unfolding pathways
by experiments.

\vskip 2 mm

\noindent{\bf Thermal unfolding pathways} \vskip 2 mm

\noindent
%{\em Thermal unfolding pathways.}
In order to study the thermal unfolding
we follow the protocol described in {\em Materials and Methods}.
Two hundreds trajectories were generated starting from the native
conformation with different random seed numbers. The fractions of
native contacts of helix A and five $\beta$-strands are averaged
over all trajectories for the time window $0 \le \delta \le 1$.
The unfolding routes are studied by monitoring these fractions as
a function of $\delta$. Above $T \approx 500$ K
the strong thermal fluctuations (entropy driven regime) make all
strands and helix A unfold almost simultaneously. Below this
temperature the statistical preference for the unfolding
sequencing is observed. We focus on $T=370$ and 425 K. As in the
case of the mechanical unfolding the cluster 2 unfolds before
cluster 1 (results not shown). However, the main departure from
the mechanical behavior is that the strong resistance to thermal
fluctuations of the cluster 1 is mainly due to the stability of
strand S2 but not of helix A (compare Fig.
\ref{cont_time_thermal_unfold_fig}{\em c} and {\em d} with Fig.
\ref{cont_ext_fig}). The unfolding of cluster 2  before cluster 1
is qualitatively consistent with the experimental
observation that
the C-terminal fragment (residues 36-76)
is largely unstructured while native-like structure persists in
the N-terminal fragment (residues 1-35)
\cite{Bofill_JMB05,Cox_JMB93,Jourdan_Biochem00}.
This is also consistent with the data from the folding simulations
\cite{Sorenson_Proteins02} as well as with the experiments of Went and Jackson
\cite{Went_PEDS05} who have shown that the $\phi$-values $\approx 0$ in the
C-terminal region. However, our finding is at odds with the high $\phi$-values 
obtained for several residues in this region by all-atom simulations \cite{Marianayagam_BPC04}
and by a semi-empirical approach \cite{Fernandez_JCP01}.
One possible reason for
high $\phi$-values
in the C-terminal region is due to the force fields.
For example, Marianayagam
and Jackson have employed the GROMOS 96 force field \cite{Gunstren_96} within
the software GROMACS software package \cite{Berendsen_CPC95}.
It would be useful to check if the other force fields give the same result or
not.

The evolution of the fraction of intra-structure contacts of A, B, C, D and E
is shown in Fig. \ref{cont_time_thermal_unfold_fig}{\em a} ($T=425$ K)
and {\em b} ($T=$370 K).
Roughly we have the unfolding sequencing,
given by Eq. \ref{thermal_sequencing_struc},
%(C,D) $\rightarrow$ E $\rightarrow$ B $\rightarrow$ A
 which strongly differs
from the mechanical one. The large stability of the $\alpha$ helix fragment
A against thermal fluctuations is consistent with the all-atom unfolding simulations
\cite{Alonso_ProSci98} and the experiments \cite{Went_PEDS05}.
The N-terminal structure B unfolds even after the core
part E and at $T=370$ K its stability is comparable with helix A.
The fact that B can withstand thermal fluctuations at high temperatures
agrees with the experimental results of Went and Jackson \cite{Went_PEDS05}
and of Cordier and Grzesiek
\cite{Cordier_JMB02} who used the notation $\beta _1/\beta _2$ instead of B.
This also agrees with the results of Gilis and Rooman \cite{Gilis_Proteins01}
who used a coarse-grained model but disagrees with results from
all-atom simulations
\cite{Alonso_ProSci98}. This disagreement is probably due to the fact that
Alonso and Daggett studied only two short trajectories and B did not
completely unfold \cite{Alonso_ProSci98}. 
The early unzipping of the structure C (Eq. \ref{thermal_sequencing_struc}) is
consistent with the MD prediction \cite{Alonso_ProSci98}.
Thus our thermal unfolding sequencing (Eq. \ref{thermal_sequencing_struc}) is more complete
compared to the all-atom simulation 
and it gives the reasonable agreement with the experiments.

We now consider the thermal unstability of individual $\beta$-strands
and helix A.
At $T$ = 370 K 
(Fig. \ref{cont_time_thermal_unfold_fig}{\em d}) the trend that S2
unfolds after S4 is more evident compared to the $T=425$ K case
(Fig. \ref{cont_time_thermal_unfold_fig}{\em c}). Overall, the
simple Go model leads to the sequencing given by Eq. \ref{thermal_sequencing}.
\begin{subequations}
\begin{equation}
{\rm (C,D)} \rightarrow {\rm E} \rightarrow {\rm B} \rightarrow {\rm A}
\label{thermal_sequencing_struc}
\end{equation} 
\begin{equation} 
{\rm S5} \rightarrow {\rm S3} \rightarrow {\rm S1} \rightarrow {\rm A}
 \rightarrow {\rm (S4,S2)}.
\label{thermal_sequencing}
\end{equation}
\end{subequations}
From Eq. \ref{mechanical_sequencing}, \ref{mechan_fixN_sequencing}
and \ref{thermal_sequencing} it is
obvious that the thermal unfolding pathways of individual strands
 markedly differ from
the mechanical ones. This is not surprising because the force should unfold
the termini first while under thermal fluctuations the most unstable part
is expected to detach first.
Interestingly, for the structures the thermal and mechanical
 pathways 
(compare Eq. \ref{thermal_sequencing_struc}
and \ref{mechan_fixN_sequencing_struc}) are almost identical except that
the sequencing of C and D is less pronounced in the former case. 
This coincidence is probably accidental.

The fact that S5 unfolds first agrees with the high-resolution NMR
data of Cordier and Grzesiek \cite{Cordier_JMB02} who studied the
temperature dependence of hydrogen bonds of Ub.
However, using the $\psi$-value analysis Krantz {\em et al} \cite{Krantz_JMB04}
have found that S5 (B3 in their notation) breaks even after S1 and S2.
One of possible reasons is that, as pointed out by Fersht \cite{Fersht_PNAS04},
if there is any plasticity in the transition state which can accomodate the
crosslink between the metal and bi-histidines, then $\psi$-values would be
significantly greater than zero even for an unstructured region, leading to
an overestimation of structure in the transition state. 
In agreement with our results,
the $\phi$-value analysis \cite{Went_PEDS05}  yields that S5 breaks before S1
and A but it fails to determine whether S5 breaks before S3.
By modeling the
amide I vibrations Chung {\em et al.} \cite{Chung_PNAS05}
 argued that S1 and S2 are
more stable than S3, S4 and S5. Eq.
\ref{thermal_sequencing} shows that the thermal stability of S1
and S2 is indeed higher than S3 and S5 but S4 may be more stable
than S1. The reason for only partial agreement between our results
and those of Chung {\em et al.} remains unclear. It may be caused
either by the simplicity of the Go model or by the model  proposed
in Ref. \cite{Chung_PNAS05}.
The relatively high stability of S4 (Eq. \ref{thermal_sequencing}) is 
supported by the $\psi$-value analysis \cite{Krantz_JMB04}.

\vskip 2 mm

\noindent{\bf Thermal unfolding barrier} \vskip 2 mm

\noindent
%{\em Thermal unfolding barrier}. 
Figure  \ref{uftime_T_fig} shows the temperature dependence of the
unfolding time $\tau_{UF}$ which depends on the thermal unfolding
barrier, $\Delta F^T_{UF}$, exponentially, $\tau_{UF} \approx
\tau_{UF}^0 \exp(\Delta F^T_{UF}/k_BT)$. From the linear fit in
Fig. \ref{uftime_T_fig} we obtain $\Delta F^T_{UF} \approx 10.48
\epsilon_h \approx 10.3$ kcal/mol.
It is interesting to note that $\Delta F^T_{UF}$ is compatible with
$\Delta H_m \approx 11.4$ kcal/mol obtained from the equilibrium data
(Fig. \ref{diagram_fN_fig}{\em b}). However, the latter is defined
by  an equilibrium constant (the free energy difference
between native and denatured states) but not by the rate constant
(see, for example, Ref. \onlinecite{Noronha_BJ04}). 

%\section{Conclusion}

\vskip 6 mm
\noindent{\Large \bf Discusion}
\vskip 2 mm

We have studied the refolding of Ub following the same protocol as in the
force-clamp experiments of Fernandez and Li \cite{Fernandez_Sci04}. Under the
low quenched force the refolding is two-stage process characterized by
two different time scales $\tau^A_1$ and $\tau^A_2$,
where
$\tau^A_1 << \tau^A_2$. This result further strengthens our previous
prediction \cite{Li_PNAS06} that the nature of the folding starting from the FDE does
not depend on details of models. The simple C$_{\alpha}$-Go model provides
reasonable estimates for the equilibrium
critical force $f_c$ as well as the averaged distance between the D
and TS states, $\Delta x_F$, and the distance between the N and TS states,
$\Delta x_{UF}$. We have also obtained
%demonstrated explicitly
$\Delta H_m$ from the two-state fit
of the population of the NBA, $f_N$, and the thermal unfolding
barrier $\Delta F^T_{UF}$. It would be interesting to measure $\Delta F^T_{UF}$
experimentally and compare it with $\Delta H_m$.

The shortcoming of the Go model we used is its failure to capture
seldom unfolding intermediates observed in the
experiments \cite{Schlierf_PNAS04}
as well as in the all-atom simulations \cite{Irback_PNAS05}.
However it mimics
the overall two-state behavior of Ub.
Our simulations suggest that the non-native interactions, neglected
in the $C_{\alpha}$ Go model, may be the cause of
mechanical unfolding intermediates.

Due to thermal fluctuations the thermal unfolding pathways are not
well defined as in the mechanical case. Nevertheless,
at $T < 500$ K the statistical
preference in the sequencing of unfolding events is evident.
In accord with the experiments, the cluster 2 unfolds before cluster 1
in the mechanical as well as in the thermal cases. However, in terms of
individual strands we predict
that mechanical and thermal unfolding follows very
different pathways (Eq. \ref{mechanical_sequencing} and
Eq. \ref{thermal_sequencing}). Mechanically strand S1 is the most unstable
whereas the thermal fluctuations break contacts of S5 first.
If we consider only breaking of intra-structure native contacts, then
our mechanical sequencing
agrees with the all-atom
simulation results \cite{Irback_PNAS05}. It is probably not unexpected
because mechanical unfolding pathways may depend largely on the topology
of the native conformation and in some cases
the Go-like models may give results comparable
with experimental ones \cite{West_BJ06}.
However, contrary to  Irb\"ack {\em et al.} \cite{Irback_PNAS05},
we predict that
the terminal strands follow the mechanical unfolding
sequencing: S1 $\rightarrow$ S2  $\rightarrow$ S5. 
It would be very exciting to perform the AFM experiments
to verify this prediction and the whole unfolding
sequencing (Eq. \ref{mechanical_sequencing})

We have considered the effect of fixing one end on unfolding
kinetics and
found that it delays the unfolding by nearly a factor of 2 regardless to
what end is anchored.
We argue that
this general result may be understood, using the diffusion-collision model
developed by
Karplus and Weaver \cite{Karplus_Nature76}. However, fixing one terminus
does not affect the distance between the native
state and TS. One of the most interesting results is
that what terminus we keep fixed  matters for the unfolding
sequencing. Namely, anchoring the N-end changes it dramatically
(see Eq. \ref{mechanical_sequencing} and Eq. \ref{mechan_fixN_sequencing})
whereas fixing the C-end has only a minor effect.

As evident from Eqs. \ref{thermal_sequencing_struc} and \ref{thermal_sequencing}
and the detailed discussion in the Introduction,
our thermal unfolding sequencing
is more complete compared with previous theoretical studies 
\cite{Alonso_ProSci98,Fernandez_JCP01,Fernandez_Proteins02,Fernandez_PhysicaA02,Gilis_Proteins01,Sorenson_Proteins02}.
We have obtained some agreement with
the experimental data \cite{Cordier_JMB02,Chung_PNAS05,Went_PEDS05,Krantz_JMB04}
on the instability of
the structures and $\beta$-strands. However the picture for 
thermal unfolding
pathways is still incomplete.
 More experiments are needed to check our prediction
given by Eqs. \ref{thermal_sequencing_struc} and \ref{thermal_sequencing}. 

We have also shown that refolding from FDE and folding from TDE
have the same pathways which are not sensitive to the quenched force.
The refolding/folding sequencing is the same as for the
thermal unfolding (see Eqs. \ref{thermal_sequencing_struc} and
 \ref{thermal_sequencing}) but in the inverse order
implying that the protein folding is the reversible process.

MSL thanks A. Irb\"ack, D. K. Klimov, S. Mitternacht,
E. P. O'Brien Jr. and D. Thirumalai
for very useful discussions.
This work was supported by
the KBN grant No 1P03B01827,
%the National Science Foundation grant
%(NSF CHE-0209340),
National Science Council in Taiwan
under grant numbers No. NSC 93-2112-M-001-027 and
95-2119-M-002-001, and Academia Sinica in Taiwan under grant
numbers AS-92-TP-A09 and AS-95-TP-A07.

\newpage

%%\bibliographystyle{biophysj}
%%\bibliography{Ubiquitin_Li}

\newpage
\vskip 20mm

\begin{table}[h]
\begin{tabular}{c|c|c|c} \hline
Force (pN)~&~S1 $\rightarrow$ S2 $\rightarrow$ S5 ($\%$) &~S5
$\rightarrow$ S1 $\rightarrow$ S2 ($\%$)& (S1,S2,S5) ($\%$) \\\hline
70 & 81 & 8 & 11\\
100 & 76 & 10 & 14\\
140 & 53 & 23 & 24\\
200 & 49 & 26 & 25\\\hline
\end{tabular}
\label{tab:SimTime_trimer}
\end{table}
\noindent {\bf TABLE 1.} Dependence of unfolding pathways on the external
 force. There are three possible scenarios: S1 $\rightarrow$
 S2 $\rightarrow$ S5, S5 $\rightarrow$ S1 $\rightarrow$ S2, and three strands
unzip almost simultaneously (S1,S2,S5). The probabilities of
observing these events are given in percentage.

\newpage

\centerline{\Large \bf Figure Captions} \vskip 5 mm

\noindent {\bf FIGURE 1.} (a) Native state conformation of ubiquitin taken from the PDB
(PDB ID: 1ubq). There are five $\beta$-strands: S1 (2-6), S2 (12-16),
S3 (41-45), S4 (48-49) and S5 (65-71), and one helix A (23-34).
(b) Structures B, C, D and E consist of pairs of strands (S1,S2),
(S1,S5), (S3,S5) and (S3,S4), respectively. In the text
we also refer to helix A
as the structure A. \vskip 5 mm

\noindent {\bf FIGURE 2.} (a) The $T-f$ phase diagram obtained by the extended
histogram method. The force is applied to termini N and C.
The color code for $1-<\chi(T,f)>$ is given on the right.
The blue color corresponds to the state in the NBA, while the red
color indicates the unfolded states. The
vertical dashed line refers to $T=0.85 T_F \approx 285$ K at which
most of simulations have been performed. (b) The temperature
dependence
 of $f_N$ (open circles) defined as the
renormalized number of native contacts (see Material and Methods).
The solid line refers to
the two-state fit to the simulation data.
The dashed line represents the experimental two-state curve with
$\Delta H_{\rm m}$ = 48.96 kcal/mol and $T_m = 332.5$K \cite{Thomas_PNAS01}.
\vskip 5 mm

\noindent {\bf FIGURE 3.} (a) The dependence of the free energy on
$Q$ for selected values of $f$ at $T=T_F$. D and N refer to
the denaturated and native state, respectively. (b) The
dependence of folding and unfolding barriers, obtained from the
free energy profiles, on $f$. The linear fits $y = 0.36 + 0.218x$
and $y=0.54 - 0.029x$ correspond to $\Delta F_{F}$ and $\Delta
F_{UF}$, respectively. From these fits we obtain $\Delta x_F
\approx $ 10 nm and $\Delta x_{UF} \approx$ 0.13 nm.
\vskip 5 mm

\noindent {\bf FIGURE 4.} (a) and (b) The time dependence of $Q$, $R$ and $R_g$
for two typical trajectories starting from FDE
($f_q=0$ and $T=285$ K).
The arrows 1, 2 and 3 in (a) correspond to time 3.1 ($R=10.9$ nm),
9.3 ($R=7.9$ nm) and 17.5 ns ($R=5$ nm).
The arrow 4 marks the folding time $\tau _F$ = 62 ns ($R=2.87$ nm)
when all of 99
native contacts are formed. (c) and (d) are the same as in (a) and (b) but
for $f_q$ = 6.25 pN. The corresponding arrows refer to $t=$
7.5 ($R=11.2$ nm), 32 ($R=9.4$ nm), 95 ns
($R=4.8$ nm) and $\tau _F = 175$ ns ($R=3.65$ nm). \vskip 5 mm

\noindent {\bf FIGURE 5.} (a) The time dependence of $<Q(t)>$, $<R(t)>$ and $<R_g(t)>$
 when the refolding starts from TDE.
(b) The same as in (a) but for the short time scale.
(c) and (d) The same as in (a) and (b) but for FDE with $f_q=0$.
(e) and (f) The same as in (c) and (d) but for $f_q$=6.25 pN.
\vskip 5 mm

\noindent {\bf FIGURE 6.} The dependence of folding times on the quench force at
$T=285$ K.
$\tau_F$ was computed as the average of the first passage times
($\tau_F$ is the same as $\tau^Q_2$ extracted from the
bi-exponential fit for $<Q(t)>$). The result is averaged over 30 -
50 trajectories depending on $f_q$. From the linear fit $y = 3.951
+ 0.267x$ with correlation level equal -0.96 we obtain $x_F
\approx$ 0.96 nm. 
\vskip 5 mm

\noindent {\bf FIGURE 7.} Time dependence of the end-to-end distance for $f=100$ pN.
The thin curves refer to fifteen representative trajectories. The averaged
over 200 trajectory $<R(t)>$ is represented by the thick line. The dashed curve
is the single exponential fit $<R(t)> = 21.08 -16.81\exp(-x/\tau_{UF})$,
where $\tau_{UF}\approx 11.8$ ns.
\vskip 5 mm

\noindent {\bf FIGURE 8.} Dependence of mechanical unfolding time on the force.
Circles refer to the process when the force is applied to both
N and C termini. Squares signifies the case when the N-end is fixed and
the C-end is pulled. For the first case  
the linear fit ($y=9.247 -0.067x$)  gives the distance
between the native state and TS $\Delta x_{UF} \approx $ 0.24 nm.
In the second case, from the linear fit
($y=9.510 -0.062x$) we obtained $\Delta x_{UF} \approx $ 0.22 nm. Thus,
within error bars fixing one end does not affect the value of $\Delta x_{UF}$.
The inset shows the linear dependence
of $\tau_{UF}$ on $f$ in the high force regime.
\vskip 5 mm

\noindent {\bf FIGURE 9.} The dependence of fraction of the native contacts on
the end-to-end extension
for cluster 1 (solid lines) and cluster 2 (dashed lines) at
$f=70 pN$  and 200 pN.
The results are averaged over 200 independent
trajectories.
The arrow points to the extension $\Delta R$ = 8.1 nm.
\vskip 5 mm

\noindent {\bf FIGURE 10.} (a) The dependence of fraction of the intra-structure native contacts
on $\Delta R$ for structures A, B, C, D and E  at
$f=100$ pN. (b) The same as in a) but for all native contacts.
The results are averaged over 200 independent
trajectories. 
\vskip 5 mm

\noindent {\bf FIGURE 11.} (a) The dependence of fraction of the native contacts on $\Delta R$
for strand S1, S2 and S5
($f=200 pN$). The vertical dashed line marks the position of the plateau
at $\Delta R \approx$ 1 nm. (b) The snapshot, chosen at the extension
marked by the arrow in a), shows that S2 unfolds before S5.
At this point all native contacts of S1 and S2 have already broken
while 50$\%$ of the native contacts of S5 are still present.
\vskip 5 mm

\noindent {\bf FIGURE 12.} (a) The dependence of fraction of the native contacts on extension
for A and all  $\beta$-strands at
$f=70 pN$. (b) The same as in (a) but for $f=200$ pN. The arrow points to
$\Delta R = 8.1$ nm where the intermediates are recorded on the experiments
\cite{Schlierf_PNAS04}. The results are averaged over 200 trajectories.
\vskip 5 mm

\noindent {\bf FIGURE 13.}
(a) The dependence of fraction of the
intra-structure native contacts on extension
for all  structures at
$f=200 pN$. The N-terminus is fixed and the external force is applied via
the C-terminus. (b) The same as in (a) but for the native contacts of all
individual $\beta$-strands and helix A .
The results are averaged over 200 trajectories.
(c) A typical snapshot which shows that S$_5$ is fully detached from
the core while S$_1$
 and S$_2$ still have $\approx 50\%$ and 100\% contacts, respectively.
(d) The same as in (b) but the C-end is anchored and N-end is pulled.
The strong drop in the fraction of native contacts of S$_4$ at $\Delta R \approx 7.5$ nm does not correspond
to the substantial change of structure as it has only 3 native
contacts in total.

\vskip 5 mm

\noindent {\bf FIGURE 14.} (a) The dependence of fraction of intra-structure
native contacts on
the progressive variable $\delta$ for all structures at $T$=425 K.
(b) The same as in (a) but for $T=370$ K. (c) The dependence of
the all native contacts of the $\beta$-strands and helix A at
$T$=425 K. (d) The same as in (c) but for $T=370$ K.
\vskip 5 mm

\noindent {\bf FIGURE 15.} Dependence of thermal unfolding time $\tau _{UF}$ on
 $\epsilon _H/T$,
where $\epsilon _H$ is the hydrogen bond energy. The straight line is a fit
 $y = -8.01 + 10.48x$.
\vskip 5 mm

\clearpage

% FIG. 1
%\centerline{\bf FIGURE 1.}

\clearpage

\vskip 30 mm
% FIGURE 2
%\centerline{\bf FIGURE 2.}

\clearpage

\vskip 30 mm

% FIGURE 3
%\centerline{\bf FIGURE 3.}

\clearpage

\vskip 30 mm

% FIGURE 4
%\centerline{\bf FIGURE 4.}

\clearpage

\vskip 30 mm

% FIGURE 5
%\centerline{\bf FIGURE 5.}

\clearpage

\vskip 30 mm
% FIGURE 6
\begin{figure}
\includegraphics[width=15cm,angle=0]{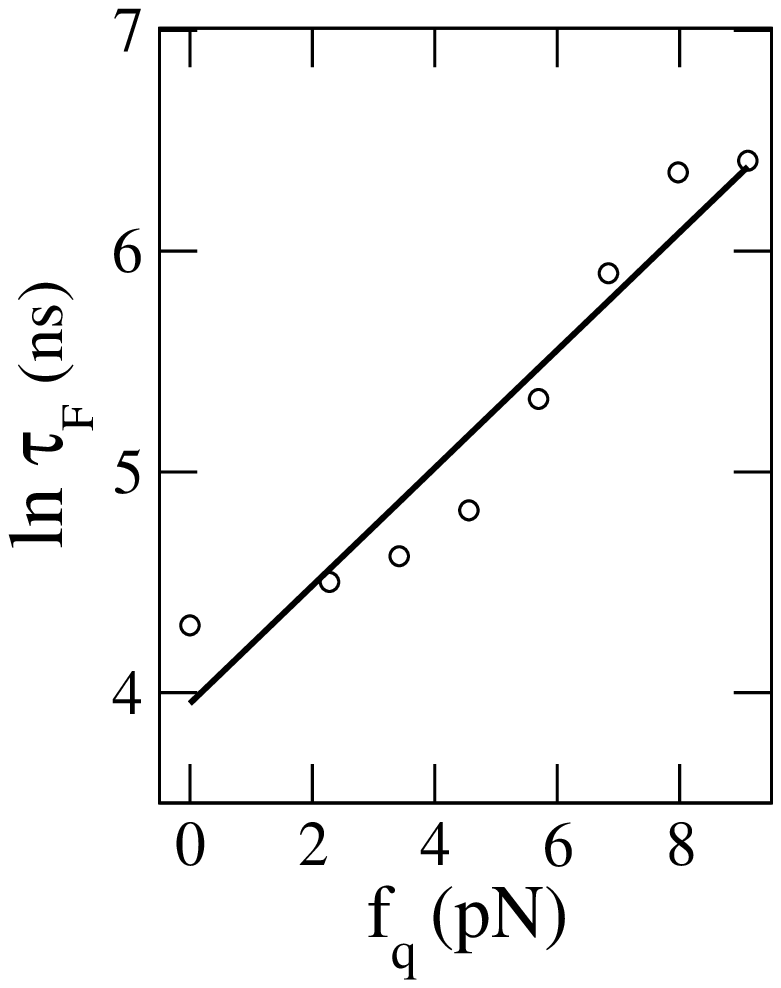}
\caption{}
\label{Kf_vs_ffc_fig}
\end{figure}
%\centerline{\bf FIGURE 6.}

\clearpage

\vskip 30 mm
% FIGURE 7
\begin{figure}
\includegraphics[width=15cm,angle=0]{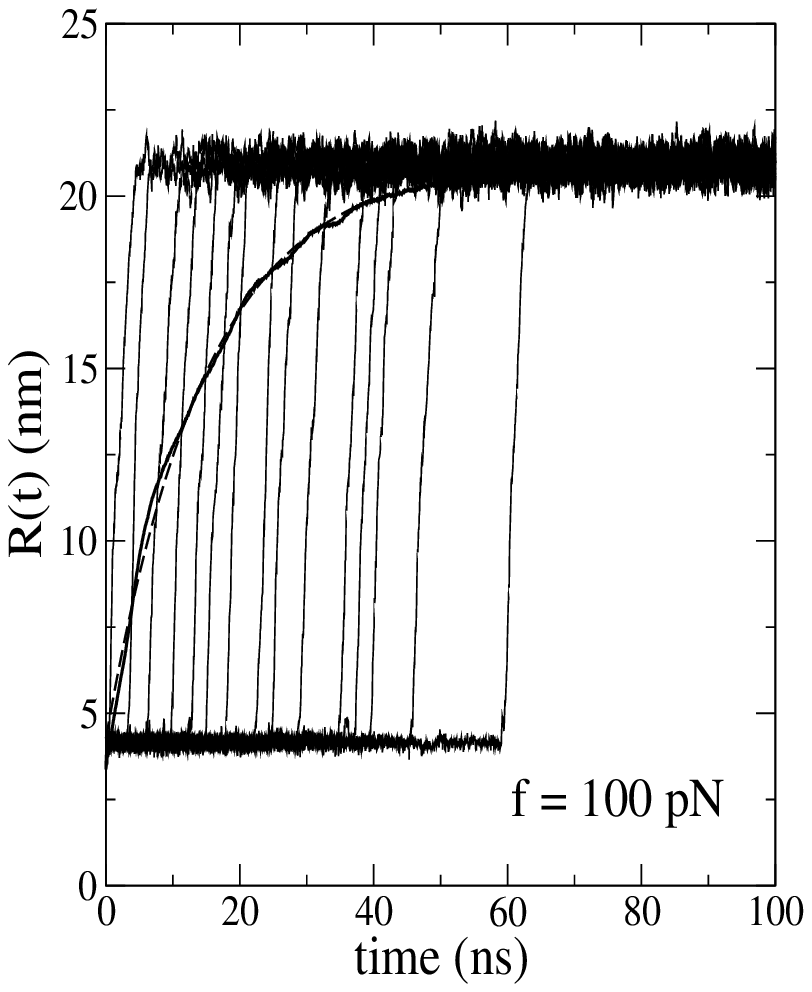}
%\vspace{0.2in}
\caption{}
\label{uf100pN_long_fig}
\end{figure}
%\centerline{\bf FIGURE 7.}

\clearpage

\vskip 30 mm
% FIGURE 8
\begin{figure}
\includegraphics[width=16cm,angle=0]{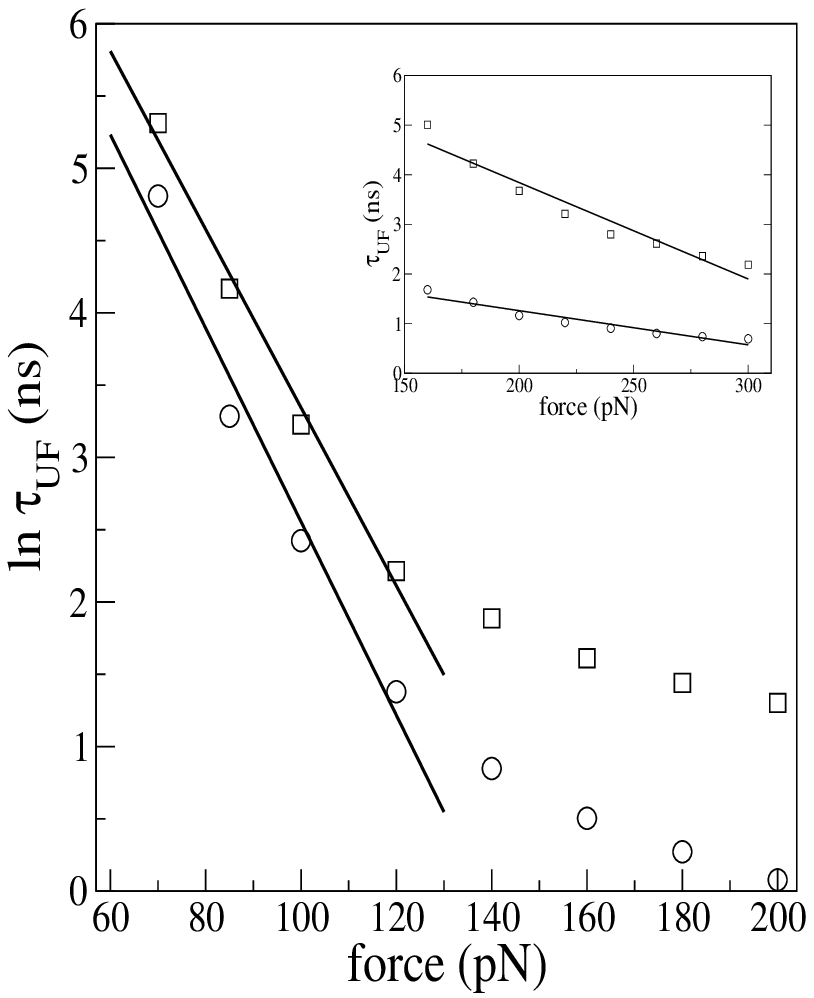}
\caption{}
\label{uftime_low_high_fig}
\end{figure}
%\centerline{\bf FIGURE 8}

\clearpage

\vskip 30 mm
% FIGURE 9
\begin{figure}
\includegraphics[width=16cm,angle=0]{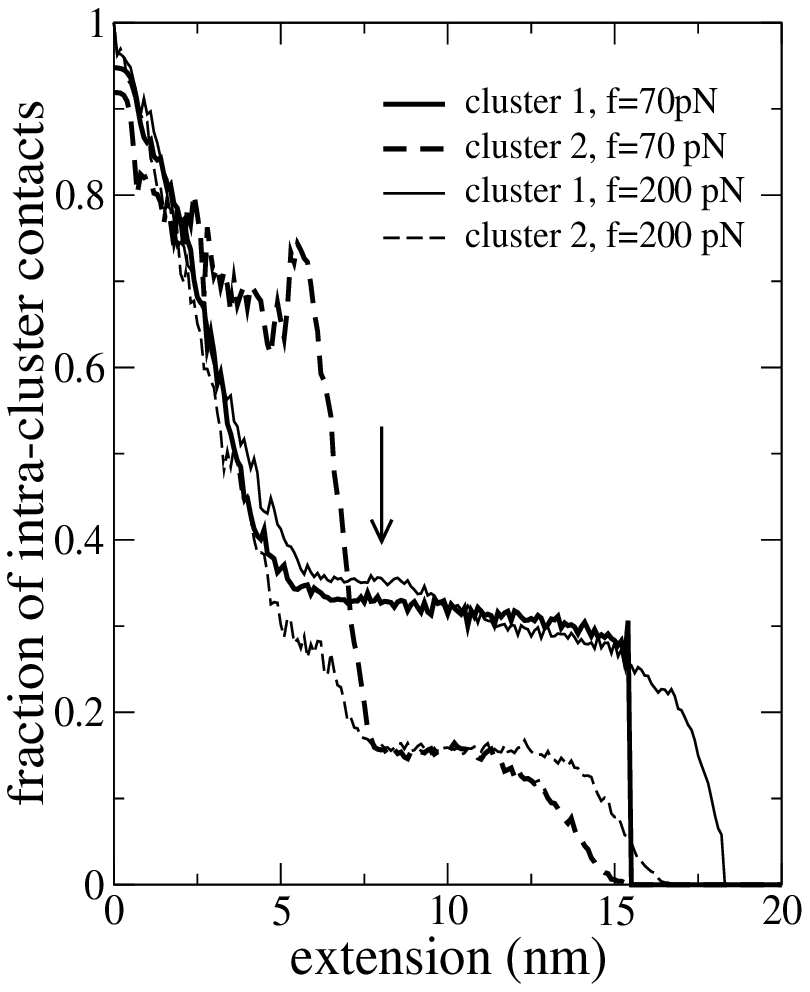}
%\vspace{0.2in}
\caption{}
\label{clus_intra_ext_fig}
\end{figure}
%\centerline{\bf FIGURE 9}

\clearpage

\vskip 30 mm
% FIGURE 10
\begin{figure}
\includegraphics[width=17cm,angle=0]{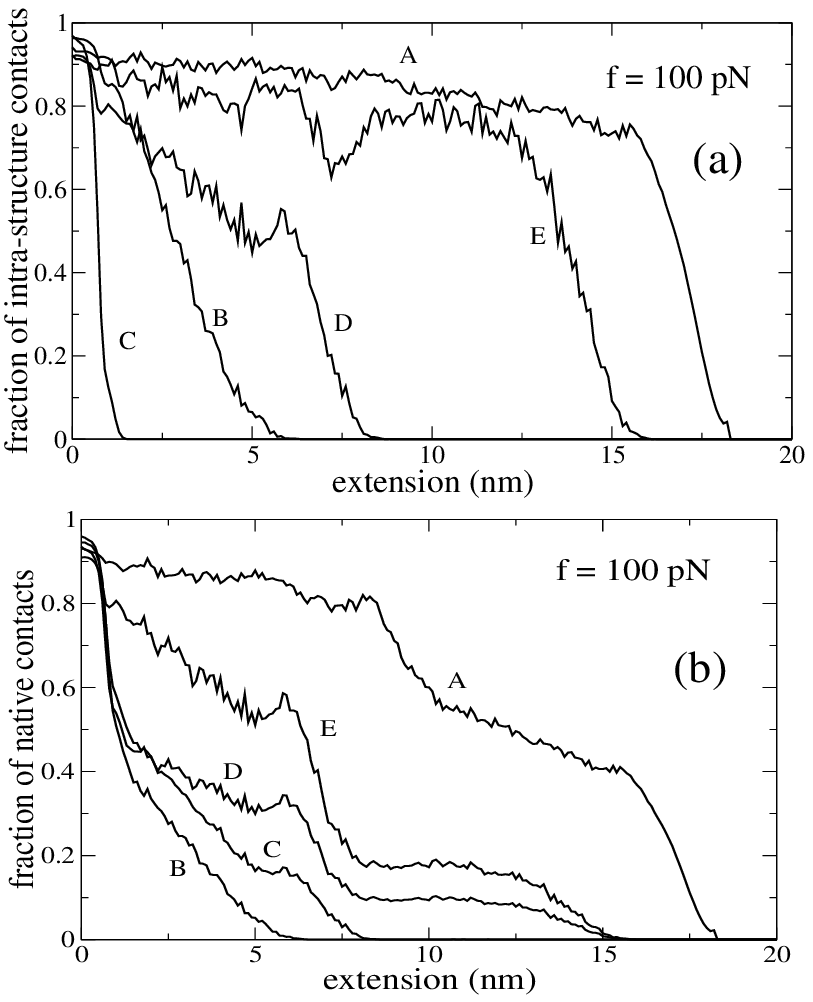}
%\vspace{0.2in}
\caption{}
\label{dom_ext_100pN_fig}
\end{figure}
%\centerline{\bf FIGURE 10}

\clearpage

\vskip 30 mm

% FIGURE 11
\begin{figure}
\includegraphics[width=18cm,angle=0]{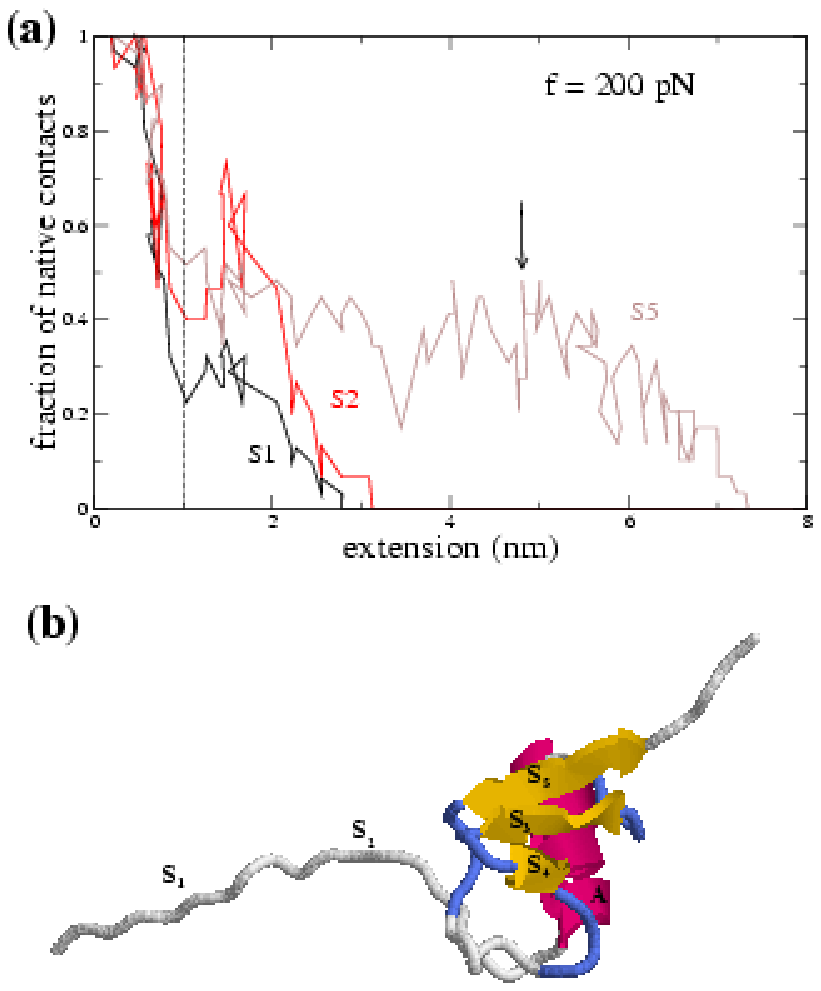}
%\vspace{0.2in}
\caption{}
\label{str_detach_cont_tr75_200pn_fig}
\end{figure}
%\centerline{\bf FIGURE 11.}

\clearpage

\vskip 30 mm

% FIGURE 12

\begin{figure}
\includegraphics[width=16cm,angle=0]{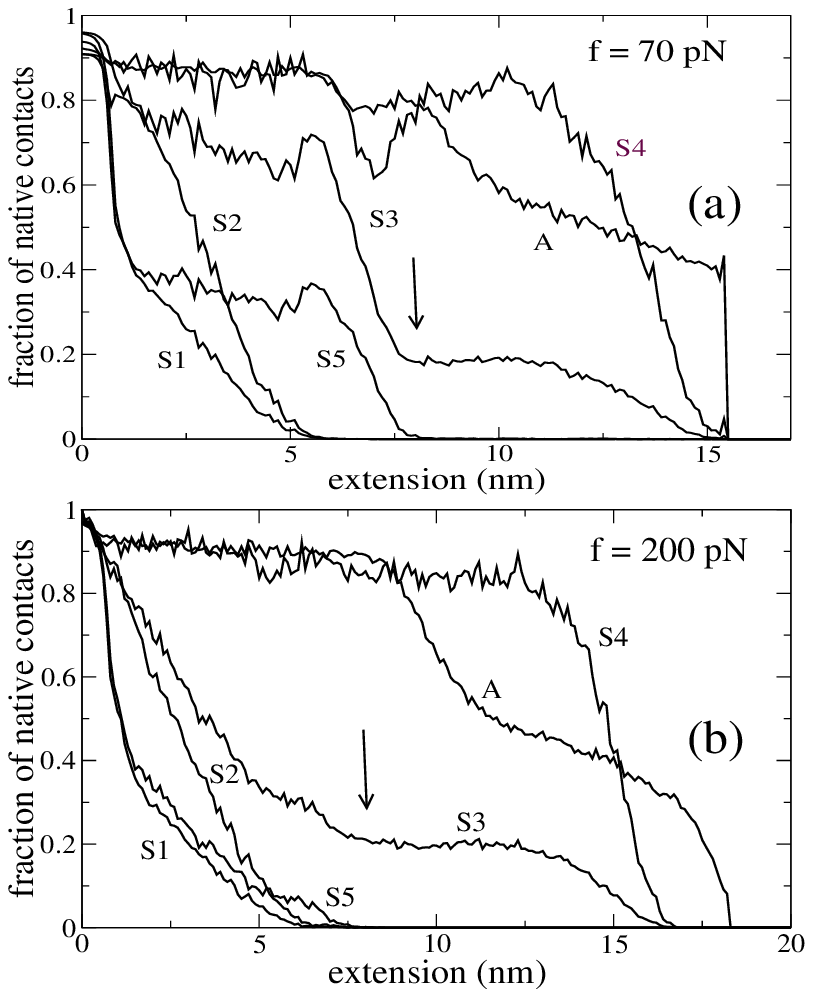}
%\vspace{0.2in}
\caption{}
\label{cont_ext_fig}
\end{figure}
%\centerline{\bf FIGURE 12.}

\clearpage

% FIGURE 13

\begin{figure}
\includegraphics[width=16cm,angle=0]{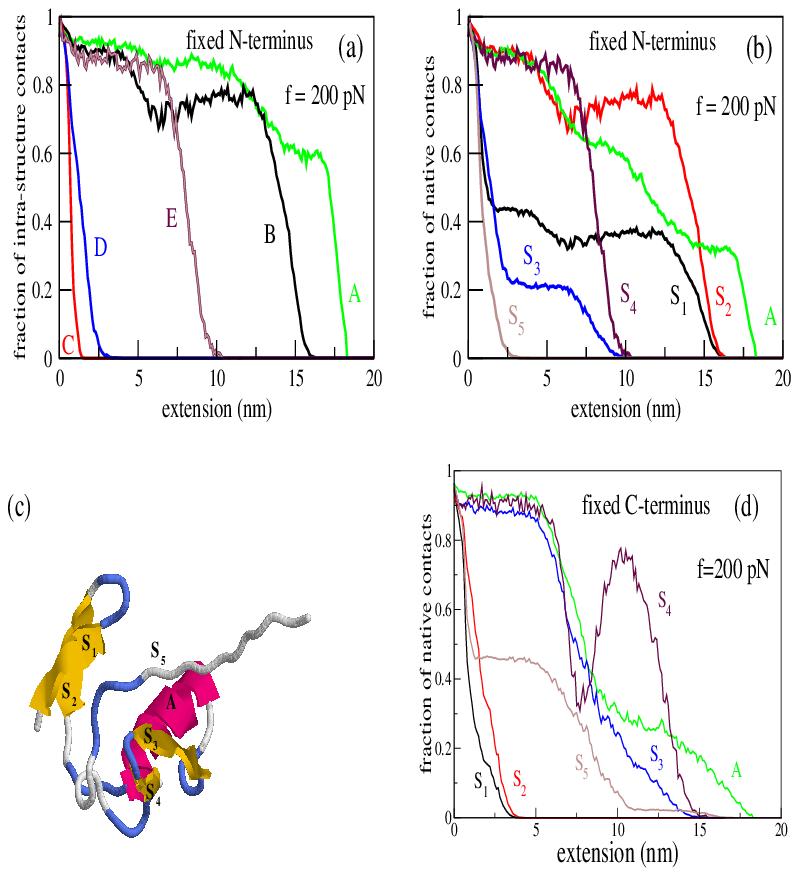}
%\vspace{0.2in}
\caption{}
\label{cont_snap_fixN_f200pN_fig}
\end{figure}

\clearpage

\vskip 30 mm

% FIGURE 14
\begin{figure}
\includegraphics[width=16cm,angle=0]{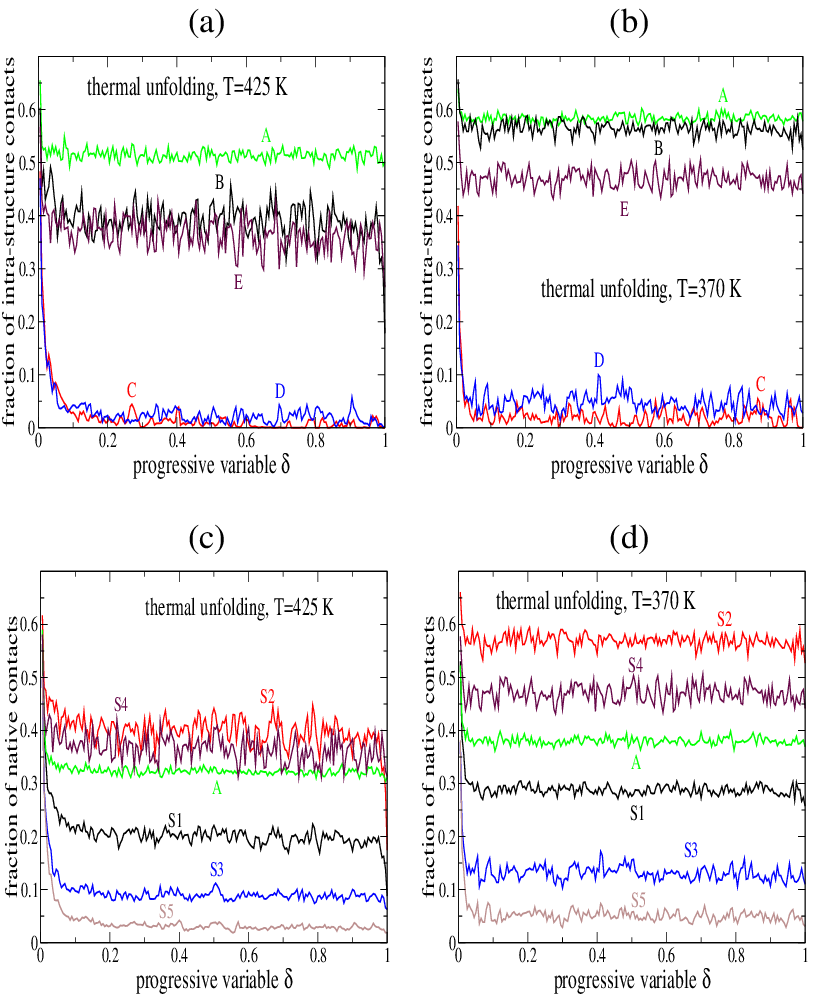}
%\vspace{0.2in}
\caption{}
\label{cont_time_thermal_unfold_fig}
\end{figure}
%\centerline{\bf FIGURE 13.}

\clearpage

\vskip 30 mm

% FIGURE 15
\begin{figure}
\includegraphics[width=16cm,angle=0]{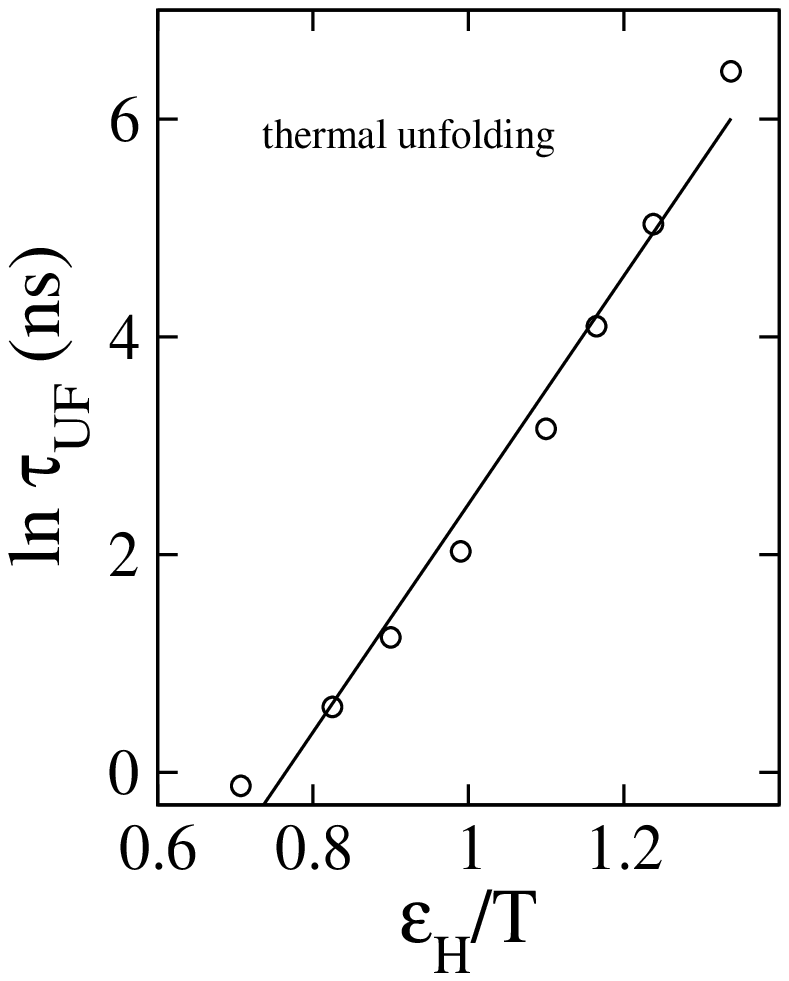}
%\vspace{0.2in}
\caption{}
\label{uftime_T_fig}
\end{figure}

\vskip 30 mm

\end{document}